\documentclass[useAMS,usenatbib]{mn2e}
\usepackage{graphicx}
\usepackage{color}
\usepackage{url}
\usepackage{longtable}
\definecolor{myred}{rgb}{0.7,0.0,0.2}

\definecolor{myblue}{rgb}{0.0,0.2,0.7}

\definecolor{mygreen}{rgb}{0.2,0.7,0.0}

\title[O star spin rates]{The spin rates of O stars in WR + O binaries. I. Motivation, methodology and first results from SALT}
\author[M. Shara et al.]{Michael M. Shara$^{1}$\thanks{E-mail: mshara@amnh.org}, Steven M. Crawford$^{2}$, Dany Vanbeveren$^{3}$, 
\newauthor {Anthony F. J. Moffat$^{4}$, David Zurek$^{1}$ and Lisa Crause$^{2}$}
\\
$^{1}$Department of Astrophysics, American Museum of Natural History, Central Park West at 79th Street, New York, NY 10024, USA\\
$^{2}$South African Astronomical Observatory, P.O. Box 9, Observatory 7935, Cape Town, South Africa\\
$^{3}$Astrophysical Institute, Vrije Universiteit Brussel, Pleinlaan 2, 1050, Brussels, Belgium\\
$^{4}$D\'epartement de Physique, Universit\'e de Montr\'eal, CP 6128 Succ. C-V, Montr\'eal, QC H3C 3J7, Canada\\
}

\begin{document}

\date{Accepted  Received }

%\pagerange{\pageref{firstpage}--\pageref{lastpage}} \pubyear{}

\maketitle

%\label{firstpage}

\begin{abstract}

The black holes (BH) in merging BH-BH binaries are likely progeny of binary O stars. Their properties, including their spins, will be strongly influenced by the evolution of their progenitor O stars. The remarkable observation that many single O stars spin very rapidly can be explained if they accreted angular momentum from a mass-transferring, O-type or Wolf-Rayet companion before that star blew up as a supernova. To test this prediction, we have measured the spin rates of eight O stars in Wolf-Rayet (WR) + O binaries, increasing the total sample size of such O stars' measured spins from two to ten.  Polarimetric and other determinations of these systems' sin i allow us to determine an average equatorial rotation velocity from HeI (HeII) lines of $v_e$ =  348 (173) km/s for these O stars, with individual star's $v_e$ from HeI (HeII) lines ranging from 482 (237) to 290 (91) km/s. We argue that the $\sim$ 100\% difference between HeI and HeII speeds is due to gravity darkening. Super-synchronous spins, now observed in all 10 O stars in WR + O binaries where it has been measured, are strong observational evidence that Roche lobe overflow mass transfer from a WR progenitor companion has played a critical role in the evolution of WR+OB binaries. While theory predicts that this mass transfer rapidly spins-up the O-type mass gainer to a nearly break-up rotational velocity $v_e \sim 530 $ km/s, the observed average $v_e$ of the O-type stars in our sample is 65\% that speed. This demonstrates that, even over the relatively short WR-phase timescale, tidal and/or other effects causing rotational spin-down must be efficient. A challenge to tidal synchronization theory is that the two longest-period binaries in our sample (with periods of 29.7 and 78.5 days) unexpectedly display super-synchronous rotation. 

\end{abstract}

\begin{keywords}
surveys -- binaries: massive -- stars: Wolf-Rayet -- stars:black holes   \end{keywords}

%==================================================================

\section{Introduction}

When massive stars rotate sufficiently rapidly (i.e. with equatorial velocities $>$ 200-300 km/s) their rotation fundamentally influences their evolution  
\citep{mae00, heg00, hir04, yoo05, bro11, eks12}. Rapid rotation may critically affect the final collapse of a massive star, leading to ultra-luminous supernovae and long-duration gamma-ray bursts \citep{woo06,geo09}. Many O-stars and their progeny, the Wolf-Rayet (WR) stars, are found in binary systems. Merging binary black holes \citep{abb16} that were generated in WR + O binaries may have spins determined by the evolution of the binary components. Those spins will be measurable with Advanced LIGO and Virgo \citep{pur16}. 

The theoretically-predicted importance of rotation has motivated several observational studies to measure the distribution of rotational velocities of massive stars. Galactic massive stars' rotation is studied in \citet{pen96, how97, vaa98}, while the VLT Tarantula survey examined similar stars in the LMC \citep{duf13, ram13}. All of these studies reached a similar conclusion: the distribution for both the early B-type stars and O-type stars is bimodal. The majority of these stars are relatively slow rotators with an average equatorial velocity $v_e \sim$100 km/s, while a smaller, but significant fraction of them are rapid rotators with $v_e > $ 200 km/s, reaching up to 500-600 km/s. The early Be-type stars obviously belong to the early B-type high velocity tail. 
 
The Geneva team (\citet{eks12} and references therein) attempted to determine the average equatorial velocity $<v_e>$ of massive single stars at birth 
 under the assumption that the overall bimodal distribution, noted above, is the norm for massive single stars in general. They concluded that $ <v_e> \sim $300 km/s. This has been the value used in most of the massive single star evolutionary calculations published by the Geneva team in the last 15 years.  
 
\subsection{Binaries} 

However, it was argued decades ago that a significant fraction of all massive stars are close binary components (for extended reviews see \citet{vdh93, van98, vaa98}), and that a significant fraction of the rapid rotators may have a binary origin. Since 1998 several extensive observational campaigns have confirmed that most (and perhaps all) massive stars are born in close binaries \citep{mas98, mas09, san11, san13}. The suggestion that ``many rapid rotators have a binary origin" has become more and more plausible. A theoretical study by \citet{dmk13}, which used overall population synthesis tools and included binaries, successfully reproduces the observed distribution of rotation rates of massive stars.

\citet{vnn98} posed the questions of whether the WR  components in WR+OB binaries formed by stellar wind mass loss, or by binary mass loss processes (Roche Lobe Overflow -RLOF - or Common Envelope -CE), and if RLOF happened in WR+OB binary progenitors, was it accompanied by mass
transfer and mass accretion? When primaries exceed $\sim$ 40-50 $M_{\odot}$ they may lose a significant fraction of their initial masses via stellar winds, greatly increasing the binary period. The existence of WR + O binaries with periods of days can only come about via RLOF or CE phases. The progenitor of a WR+OB binary, where the OB star has luminosity class V, or where the O component is an early O-type star, most likely underwent RLOF
and mass transfer, causing the rejuvenation of the gainer \citep{vnn98}.

Rapid rotation in close binaries can arise during Roche lobe overflow (RLOF) when mass lost by the mass donor is accreted by the mass gainer \citep{van98}. Mass accretion is accompanied by angular momentum accretion, hence, in this scenario, the mass gainer spins up and becomes a rapid rotator \citep{pac81}. When the mass donor ends its life with an asymmetric supernova explosion, the binary may be disrupted. This binary supernova scenario successfully predicts that the escaped component (the mass gainer) will be a rapidly rotating single runaway star (defined as a star with peculiar space velocities $>$ 30 km/s \citep{bla61}). Most of the known massive runaway stars are rapid rotators, with $\zeta$ Pup being the prototype \citep{van12}. $\zeta$ Pup displays spectral class O4 I(n)fp \citep{sot14}. While uncertainties in its distance \citep{mai08,sch08} translate directly into uncertainties in its radius (14 - 26 $R_{\odot}$) and mass (22 -56 $M_{\odot}$), and hence its critical rotation speed (550 -640 km/s) , its observed $v_e$ sin(i) of 220 km/s \citep{van12} is $\sim 35 -40\%$ of its critical rotation speed. 

\subsection{Testing a prediction of RLOF}

The motivation of this and subsequent papers is to test the prediction of RLOF-driven spin-up in massive, Wolf-Rayet (WR) + O binary stars \citep{pet05}. Most classical Wolf-Rayet (WR) stars are core helium burning objects that have lost their hydrogen-rich layers. This has occurred via stellar wind mass loss, if the WR star is a single star or in a wide, non-interacting binary, and/or by RLOF when the WR star is a close-binary component. O-type companions to WR stars that are spinning super-synchronously (with $v_e$ sin i typically $>$ 100km/s in binaries with periods shorter than about 10 days) are strongly indicative of RLOF mass and angular momentum accretion \citep{van98}. An example is WR 127, which contains a 23.9 $M_{\odot}$ O V star and a 13.4 $M_{\odot}$ WR star in a 9.555 day orbit that is inclined at 55.3 degrees to the line-of-sight \citep{del11}. The \citet{mar05} calibration for O-star radii (their table 1) yields 8.11 $R_{\odot}$ as the radius of the O star in WR 127. If the O star were rotating synchronously it would display a $v_e$ sin i of 36 km/s. The observed value is $\sim$250 km/s \citep{mas81}, which is highly super-synchronous.

A comprehensive study of the rotational velocities of the O-type components of WR+O binaries would be a valuable test of the idea that many rapid rotators have a binary origin, and have undergone RLOF mass transfer.  A literature search reveals that the $v_e $sin i value of the O-type components of WR + O binaries have been measured for only two systems, and estimated for one more. In one carefully measured case (V444 Cyg = WR139 \citep{mar94}), the O-type component is observed to be a rapid rotator, with $v_e$ sin i = 215 $ \pm 13$ km/s. In the second measured case (the brightest and nearest WR+O binary in the sky = WR 11 \citep{baa90}), the O7.5 giant displays $v_e$ sin i = 220 $\pm 20$ km/s. For HD 186943 = WR127 \citep{mas81}, the widths of the O stars' absorption lines were estimated to correspond to $v_e$ sin i $\sim 200-250 $km/s. $v_e$ sin i = 200 - 250 km/s is significantly faster than the $v_e$ corresponding to binary synchronicity in all three systems, and suggestive, though not conclusive evidence that RLOF has been important in the systems' evolution. Despite these encouraging early works, a systematic, quantitative study of O-star spins in WR + O binaries, both from a theoretical and an observational point of view, has not yet been carried out. 
 
The test of whether O companions in WR+O binaries are spun up by binary accretion can best be accomplished by measuring the observational line width parameter $v.$ sin i of the O component from line broadening of helium lines \citep{ram13, ram15}. Once this is available then one wants, ideally, to extract $v_e$, the equatorial rotation speed of the star. A binary's sin i may be obtained from polarimetry, photospheric or atmospheric eclipses, colliding-wind analysis, by assuming a mass for the O star from its spectral type if not otherwise known, or from visual binary orbits \citep{mof08}. If one assumes alignment of the spin axes of the binary components with the binary axis, then sin i is also known for the O-star axis. However, this may not always be a good assumption (e.g. \citet{vil05, vil06} for CQ Cep and CX Cep, two very-short period Galactic WR+O binaries, where the orbital and spin axes are misaligned). Therefore, one needs a large number of systems from which one can extract statistical information from $v.$ sin i to deduce $<v_e>$. This is precisely the methodology adopted in studies of rotation speeds of single field O stars \citep{ram13}, and in studies of O + O star binaries \citep{ram15}. In this study we have succeeded in measuring the $v.$ sin i of eight O stars in WR + O binaries, thereby raising the sample size with well-measured spin rates from two to ten.

%In principle one could get an estimate of ve sin i from one ÒsnapshotÓ observation of a binary WR+O system. However, often there are complications such as phase-dependent line-distortion %effects due to heating or tidal effects from the primary star.  To mitigate against this, it is essential to obtain complete and uniform orbital phase coverage (at least 10 observations for each %system, with densest time coverage near periastron passage in elliptical systems) in the binary orbit.  Once this is done, one can extract a reliable estimate of the true intrinsic ve sin i for the %O companion. In the case of the longest-period systems (> 100d), the Roche lobe overflow is accompanied by a common envelope phase where most of the mass lost by the donor will leave %the binary and significant mass accretion is not expected. A rapidly rotating O-type component is therefore not expected either in these binaries. To check this we will perform some snap-shot %observations of long-period WR binaries.

In Section 2 we describe the data and their reductions. The high resolution spectra of the helium lines of the O stars we study are presented in Section 3, as well as these stars' derived rotation rates. In section 4 we discuss the implications of our results for the overall evolution of rotational velocities in massive binaries, and we briefly summarize our results in Section 5.

\section{Observations and Data Reductions}\label{obs}
\subsection{Observations}

Observations\footnote{Observations were taken under SALT Proposal Code:
2015-1-SCI-064.}  of the target stars were obtained with the High Resolution
Spectrograph \citep{cra14} (HRS) of the Southern African Large
Telescope (SALT). The HRS is a dual beam, fiber-fed
echelle spectrograph. It was used in low resolution mode with
a 2.23" arcsec diameter fiber to provide a spectrum in
the blue arm over the spectral range of 3700-5500 \AA.  All of the spectra 
have a resolving power R $\sim$12,700 and signal-to-noise ratio S/N per pixel $>$ 150, 
designed to counterbalance the diluting effects of the accompanying WR spectrum. The
observations from the blue arm were read out by a single amplifier
with $1\times1$ binning.  All observations of our target stars were
obtained between May and June 2015 with most of the objects observed twice,
i.e. once on each of two separate nights.  During this time, a single ThAr arc and
spectral flats were obtained in this mode for the purposes of
calibration. In addition, observations of $\tau$ Sco were obtained on 7
August 2015 to provide a measurement of a source with a known, low value
for $v_e$ sin i.  Basic CCD reductions including bias subtractions, gain
corrections, and flat fielding were handled by the \texttt{ccdproc}
package \citep{cra15}.

Spectroscopic reductions of the data were carried out using the
\texttt{pyhrs} package (Crawford 2015). The software created an order
frame from a flat field image that assigned each pixel in the image to
a specific order.  Prior to extracting, each order was corrected for
spatial and spectral curvature.  A second order polynomial was fit to
the overall shape of the order, which was then removed.  In addition,
each row of the order was corrected for a small offset in the vertical
direction between each of the rows of the order.  This offset was well
below the size of the resolution element for the low resolution mode.
To calculate the wavelength solution, a spectrum was extracted by
summing the rows in the order. The resultant spectrum was then passed
to the \texttt{specidentify} task in the PySALT reduction package
(Crawford et al 2010) for wavelength identification and
calculation of the wavelength solution. Next, the target spectrum was
extracted from our observations. The extraction of the order was
performed as described with the final spectrum being a result of a
flux-weighted co-addition of all of the illuminated rows in the order
and the wavelength derived from the solution calculated for that
order. As a minimum of two exposures were taken for each
observation, the final step combined the extracted spectra from these
two exposures to produce the final spectrum for each star. HRS does
have a sky and target fiber, but as the target fiber only was used in
our analysis, all steps were only performed on the target fiber.

\subsection{Resolution of the SALT HRS}

The expected FWHM for an unresolved line at $\lambda=4750 \ \AA$ is $\sim 0.37 \ \AA$, corresponding to  a resolving power R $\sim$12,700,
based on the predicted performance of the HRS spectrograph (\citet{cra14}, Table 2). This corresponds to 
a velocity resolution of 24 km/s. As this is one of the first papers reporting results with the newly commissioned SALT HRS, and 
measured line widths are central to our analyses, we have measured the FWHM of several arc lines close to the wavelengths of the strongest HeI and HeII O stars' lines. An example is given in Figure 1, where an arc line at $\lambda 4541$ displays a FWHM of $0.30\pm0.02 \AA$, better than the predicted value.

As a further test of our methodology and the resolving power of HRS, 
we observed the bright B0.2V star $\tau$ Sco = HD 149438. 
The most recently measured $v_e$ sin i of $\tau$ Sco = 4 km/s \citep{nie12}. 
This extremely low value of $v_e$ sin i demonstrates that the
effects of thermal Doppler broadening, Stark effect, and micro and macro-turbulence are very small in a 
star of spectral type nearly identical to that of our coolest observed O star. Of course, this is no guarantee
that the O stars' He lines in the O + WR binaries that we study below do not undergo line broadening 
from one or more of the above mechanisms. Indeed, \citet{sun13} describe two magnetic O stars with rotation periods longer than one year, and hence 
with $v_e$ sin i $ \leq $ 1 km/s, whose line widths imply $v_e$ sin i $\sim$ 40-50 km/s. They note that the severe overestimates based on line widths alone are 
most likely caused by an insufficient treatment of microturbulence and macroturbulence, and that these broadening mechanisms can dominate line profiles in stars with $v_e$ sin i up to $\sim$ 40-50 km/s.

Due to structure in the outer profiles of each line preventing the lines from being well fit by a
single Gaussian, we measured the FWHM directly for each of the HeI$\lambda 4387$,
$4713$, and $4922$ lines in $\tau$ Sco (see Figure 2). The average $v.$ sin i from the three lines,
using the same methodology described below for our program stars, was $24\pm 3 $ km/s.
This observed value is, of course, instrumental, as the HRS low resolution mode broadens the very narrow lines of $\tau$ Sco to its resolution limit.
All of our program stars, discussed below, display FWHM of He absorption lines that are 10 to 20 times larger than the resolution limit of SALT's
HRS low resolution mode, corresponding to $v.$ sin i  (for HeI lines) that always exceed 200 km/s. Simulations show that the blending 
of the satellite lines seen in Figure 2 with the $4922$ line due to spin speeds in excess of 150 km/s {\it and} micro- or macroturbulence of 50 km/s
cause us to overestimate $v.$ sin i by less than 15\% in all cases.

\subsection{Line Analysis}

To determine the value for $v.$ sin i for the stars in our sample, we
followed the process as outlined in detail by \citet{ram13} and \citet{ram15}. 
This involved measuring the broadening of these stars' HeI and HeII lines 
via the full width at half maximum (FWHM). To measure the FWHM of the lines in our stars, we used the
\texttt{astropy} modeling software \citep{ast13} to fit Gaussian curves to the HeI $\lambda 4922$ and/or
HeII $\lambda 4541$ lines (WR30, WR47, and WR79 only exhibited the HeII $\lambda 4541$ line). 
Prior to fitting the Gaussian curve, a low-order polynomial was fit to the area outside of the line and
divided through to normalize the continuum. Errors for the fits were
calculated through a bootstrap Monte Carlo method of resampling the
spectra using the flux errors and repeating the fit.  From the fits,
we determined the FWHM for each of the lines and then converted the values 
for the FWHM to velocities based on the relationships in Table 1 of \cite{ram15}.

Finally, we directly measured the FWHM of the He I$\lambda4471$ line for WR 127 
from Figure 8 of \citet{del11}. 

\section{Results: The O stars' helium lines}

The fit to each of the continuum-divided absorption lines of HeI and HeII
in each of our eight WR + O binaries is presented in Figures 3 through 10. 

\newpage

\begin{figure}
\centerline{\includegraphics[width=0.99\columnwidth]{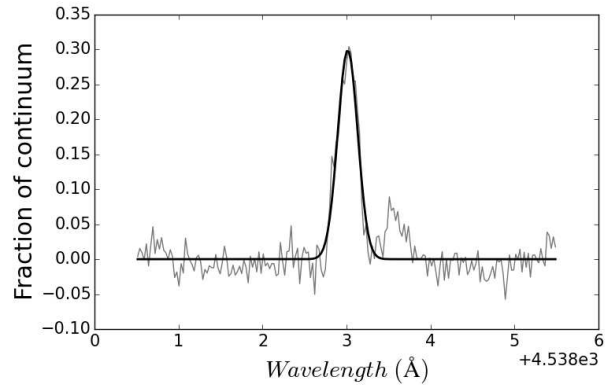}}
\caption{A SALT High Resolution Spectrograph arc line close to the HeII 4541 line. The line's FWHM is $0.30\pm0.02 \AA$. 
The data points are in light gray while the model fit is the solid black curve in this and all following figures.  }\label{spectra}
\end{figure}

\begin{figure}
\centerline{\includegraphics[width=0.99\columnwidth]{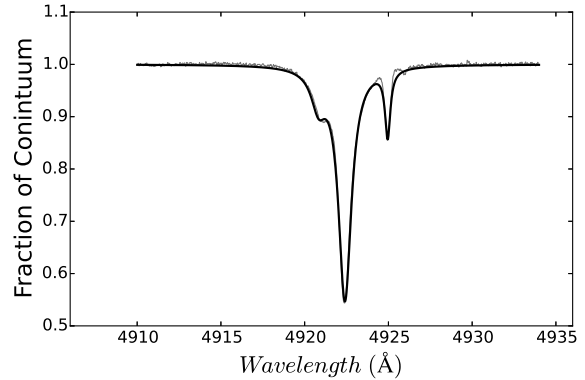}}
\caption{The HeI 4922 absorption line of the slowly rotating star $\tau$ Sco on 07 Aug 2015.}\label{spectra}
\end{figure}

\begin{figure}
\centerline{\includegraphics[width=0.99\columnwidth]{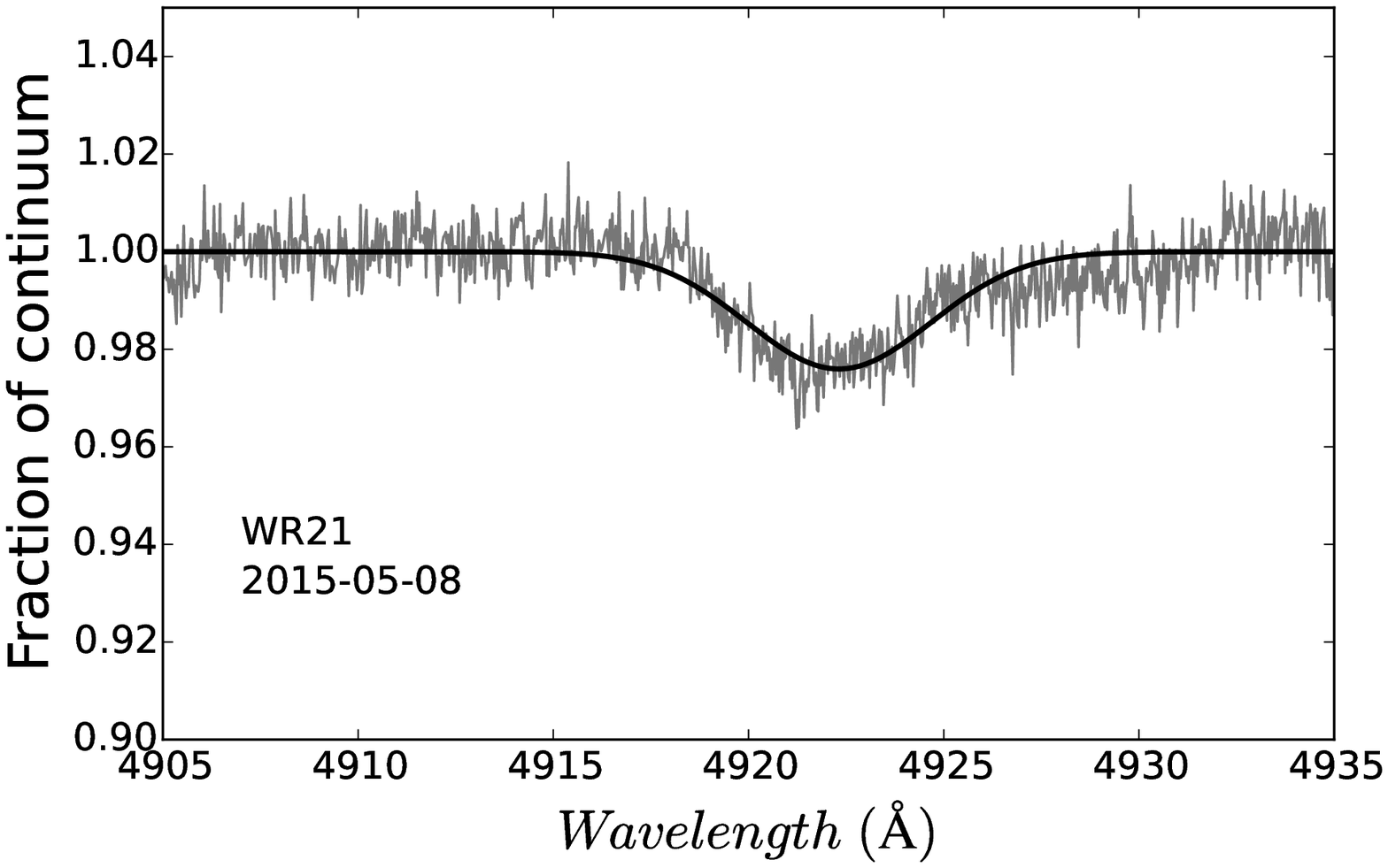}}
\centerline{\includegraphics[width=0.99\columnwidth]{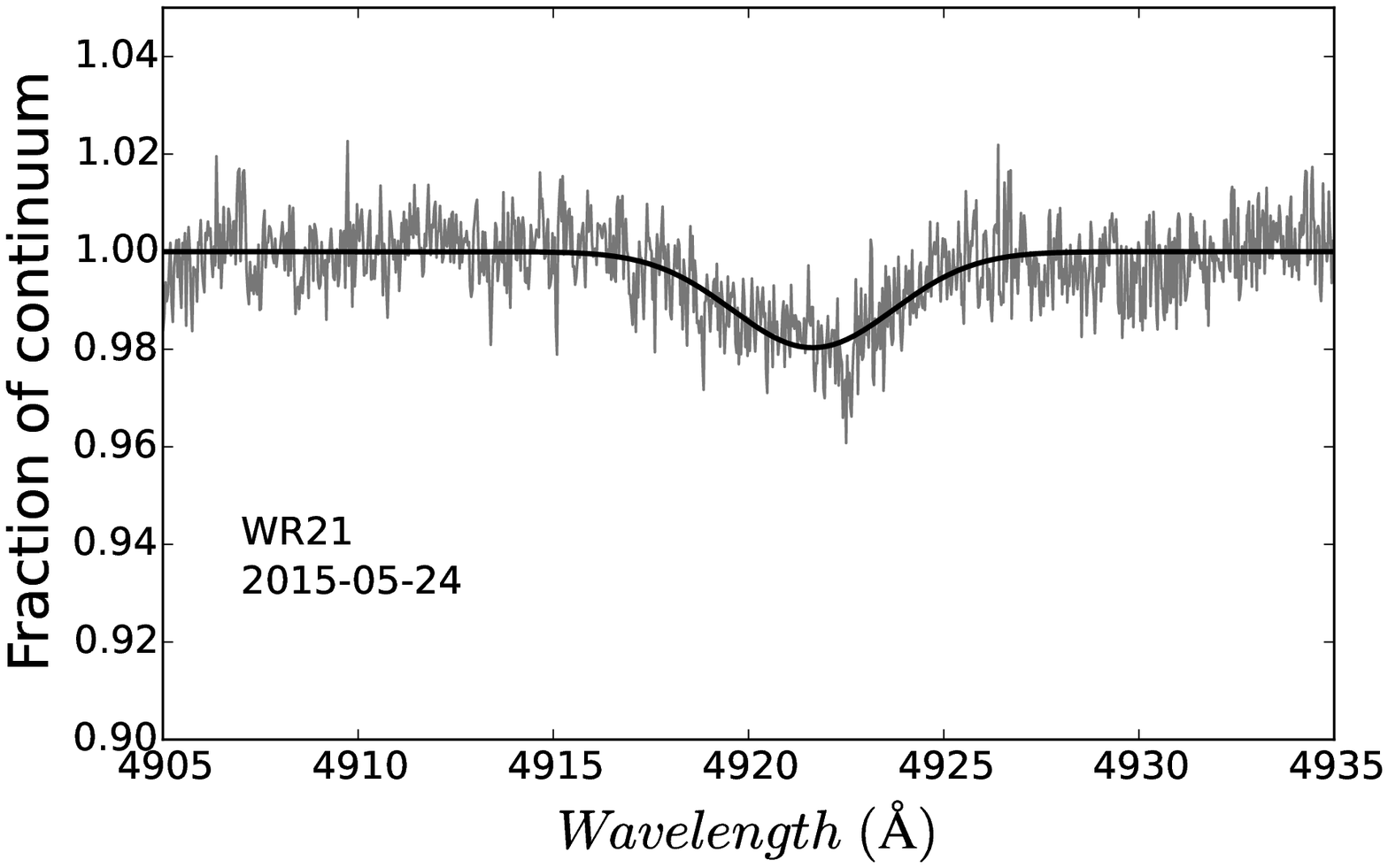}}
\centerline{\includegraphics[width=0.99\columnwidth]{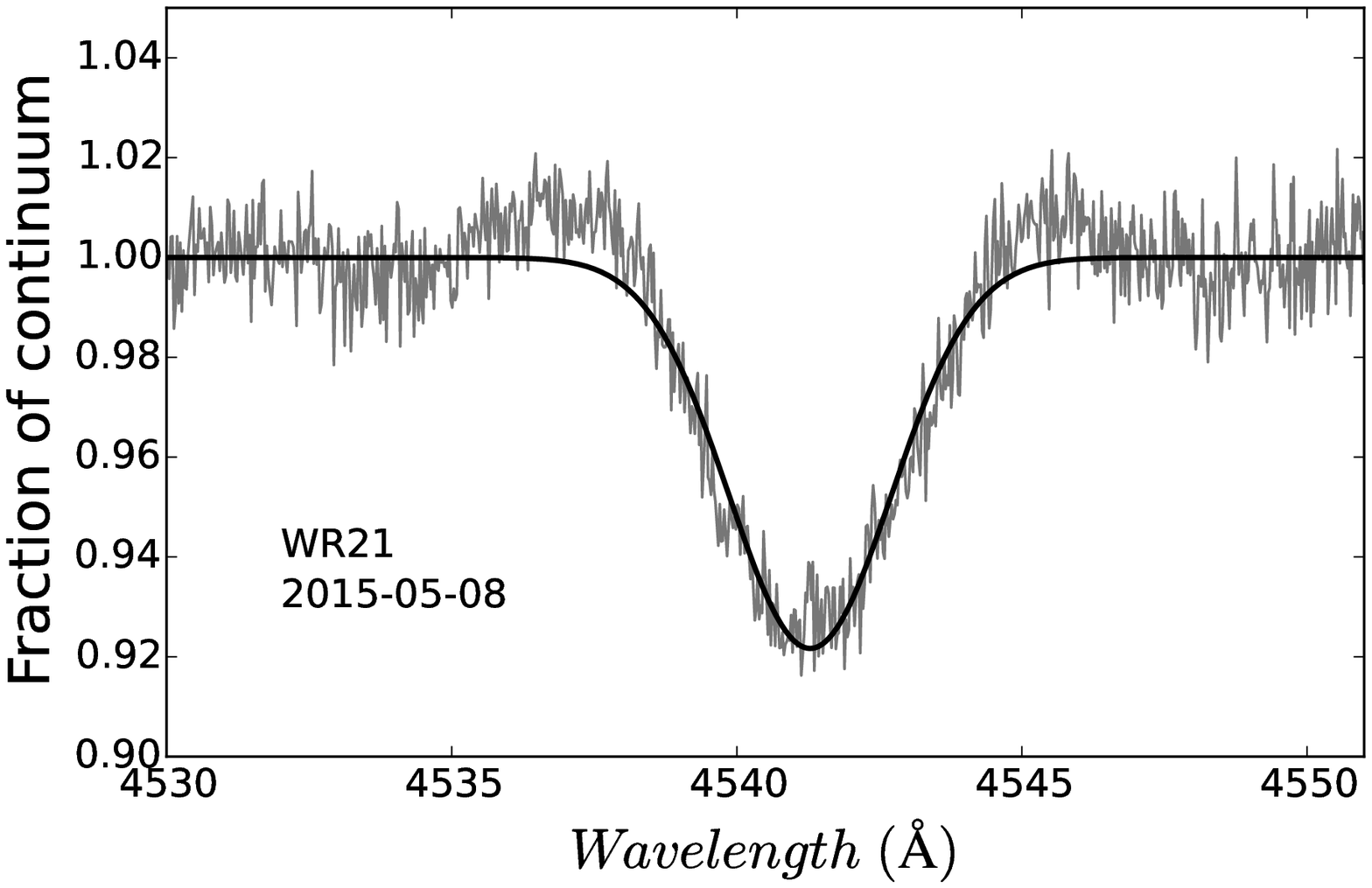}}
\centerline{\includegraphics[width=0.99\columnwidth]{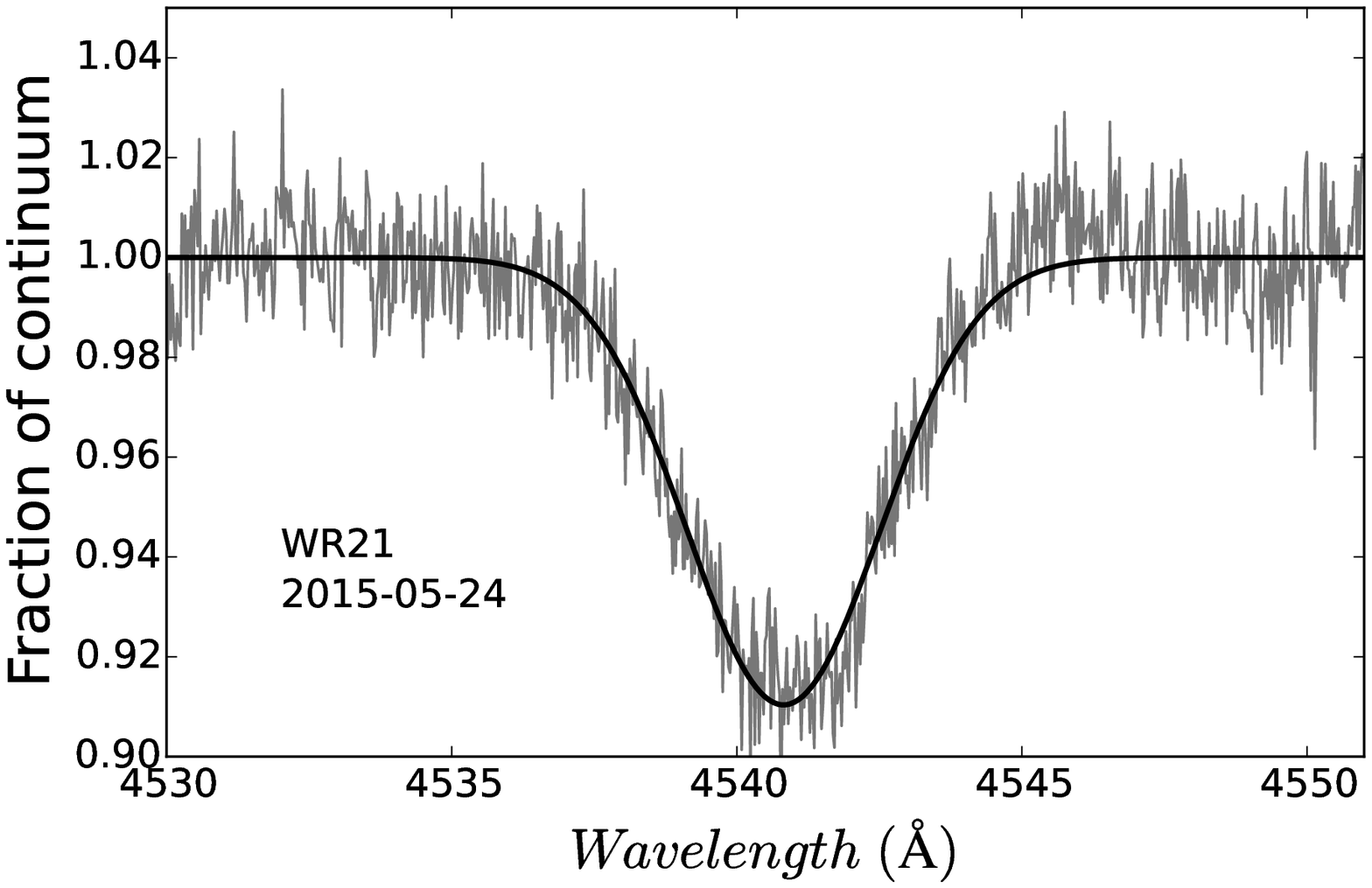}}
\caption{(Top) The HeI 4922 absorption line of WR 21 on 08 May 2015. 
(Second from Top) Same as above but on 24 May 2015.
(Third from Top) The HeII 4541 absorption line of WR 21 on 08 May 2015.
(Bottom) The HeII 4541 absorption line of WR 21 on 24 May 2015 }\label{spectra}
\end{figure}

\begin{figure}
\centerline{\includegraphics[width=0.99\columnwidth]{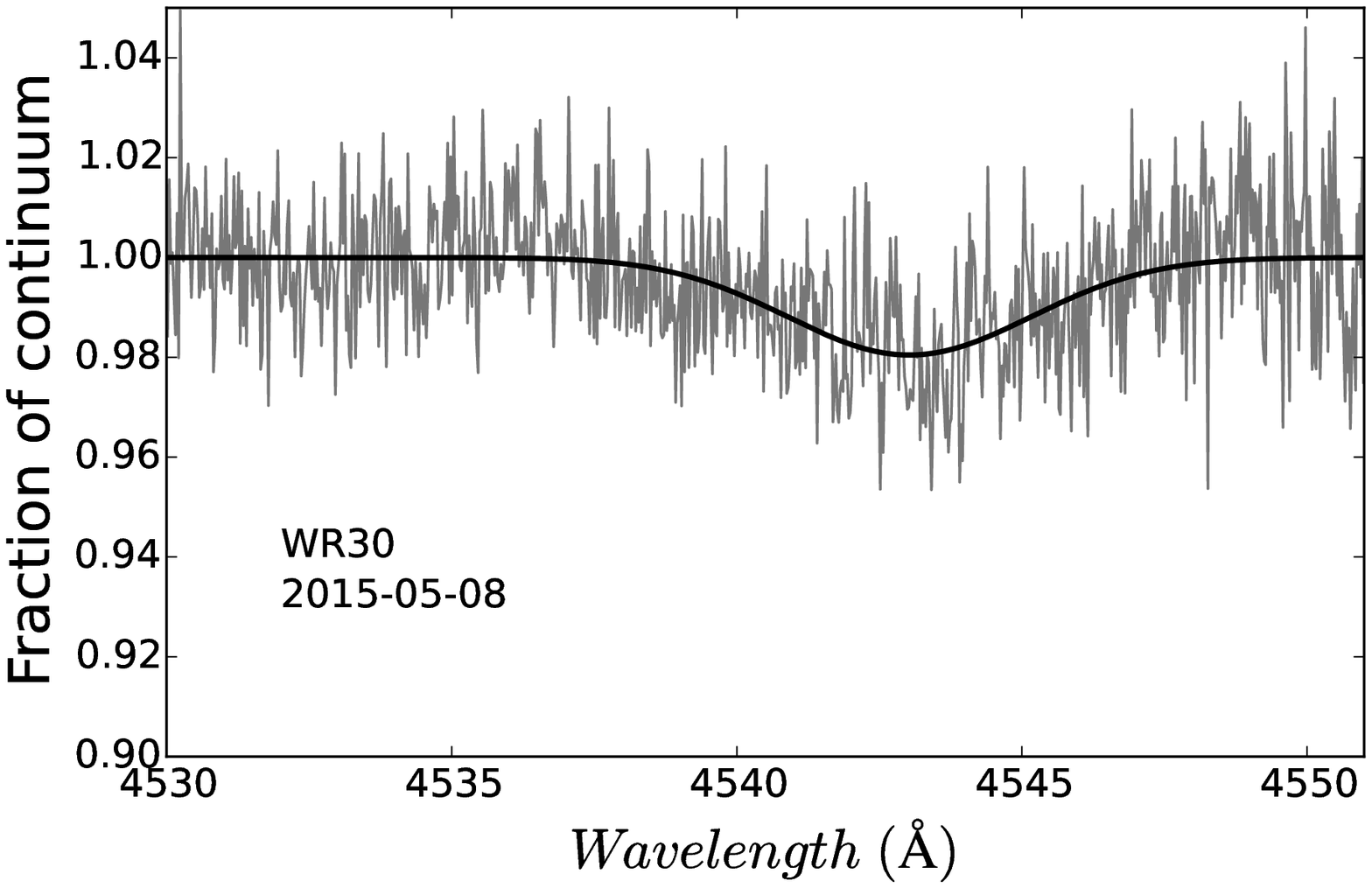}}
\caption{The HeII 4541 absorption line of WR 30 on 08 May 2015.}\label{spectra}
\end{figure}

\begin{figure}
\centerline{\includegraphics[width=0.99\columnwidth]{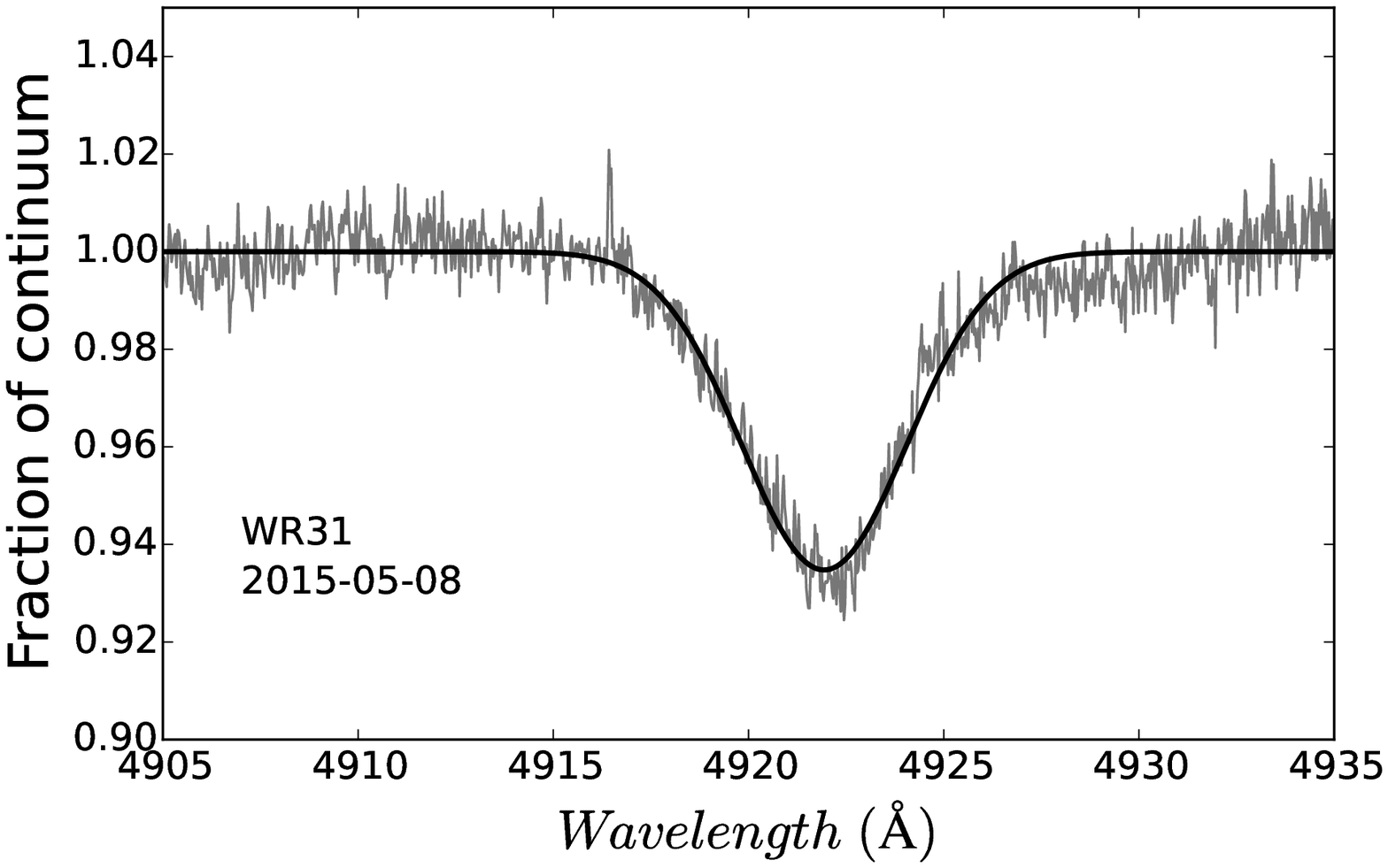}}
\centerline{\includegraphics[width=0.99\columnwidth]{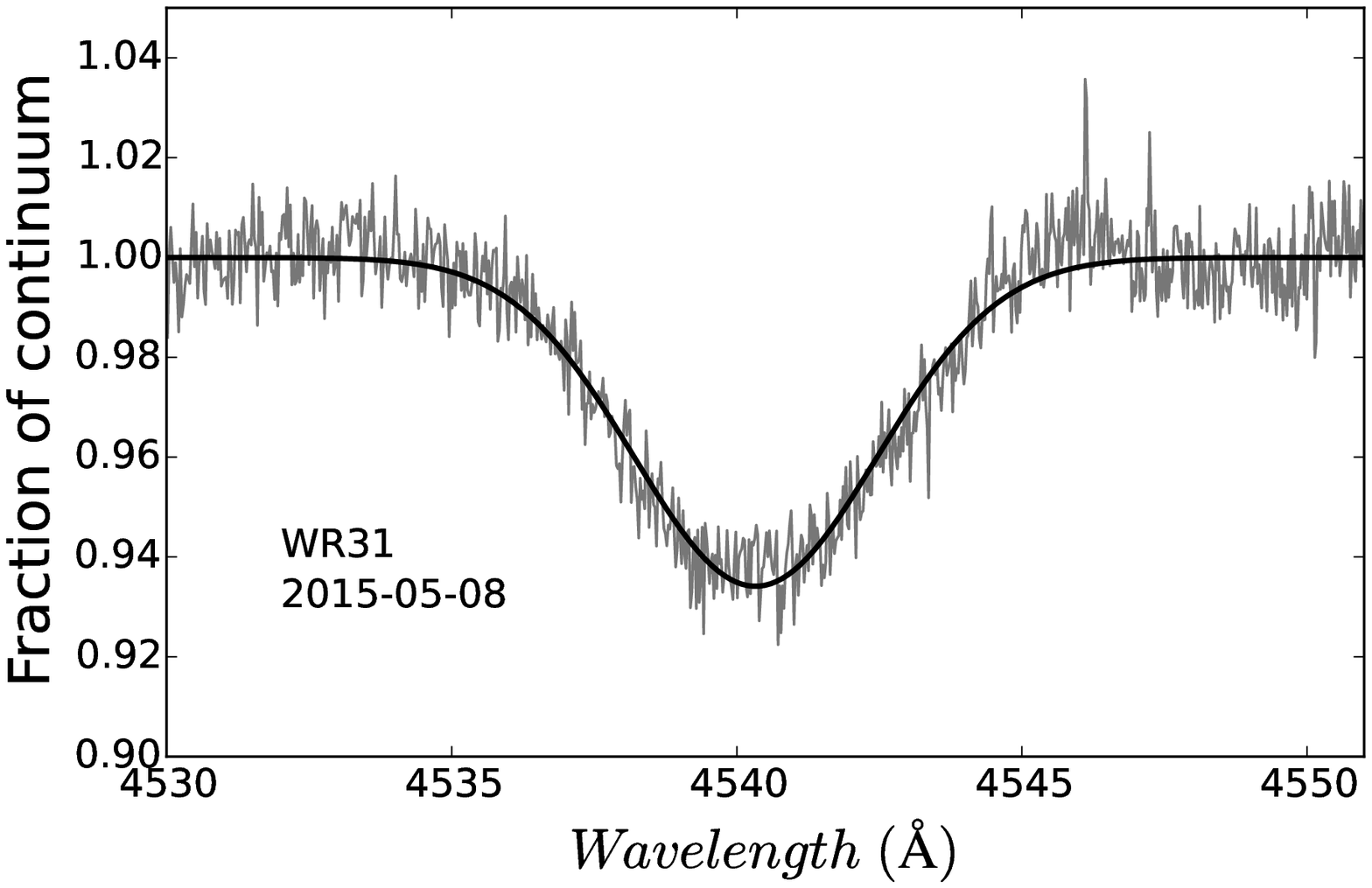}}
\caption{(Top) The HeI 4922 absorption line of WR 31 on 08 May 2015.
(Bottom) The HeII 4541 absorption line of WR 31 on 08 May 2015.}\label{spectra}
\end{figure}

\begin{figure}
\centerline{\includegraphics[width=0.99\columnwidth]{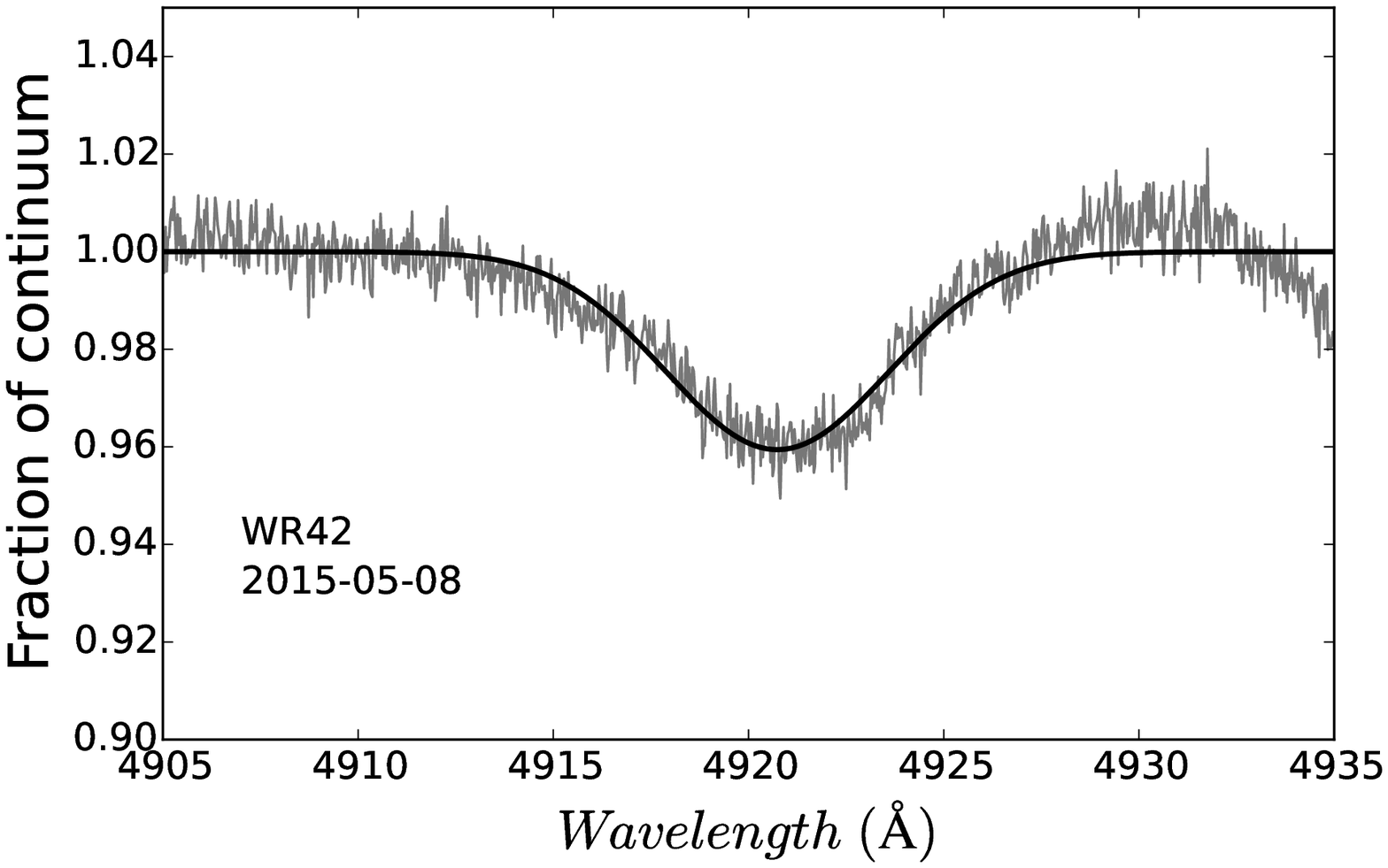}}
\centerline{\includegraphics[width=0.99\columnwidth]{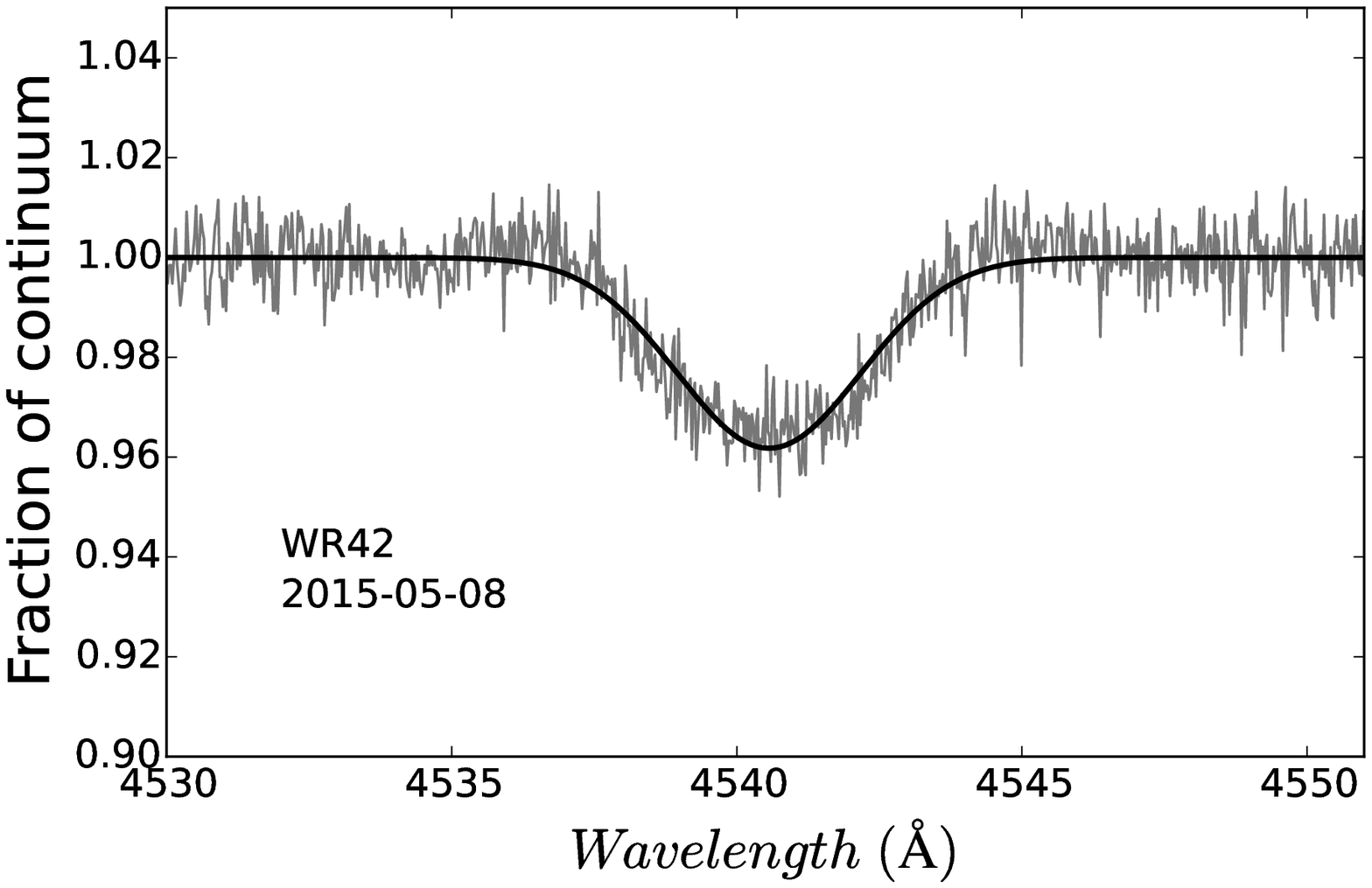}}
\caption{(Top) The HeI 4922 absorption line of WR 42 on 08 May 2015.
(Bottom) The HeII 4541 absorption line of WR 42 on 08 May 2015.}\label{spectra}
\end{figure}

\begin{figure}
\centerline{\includegraphics[width=0.99\columnwidth]{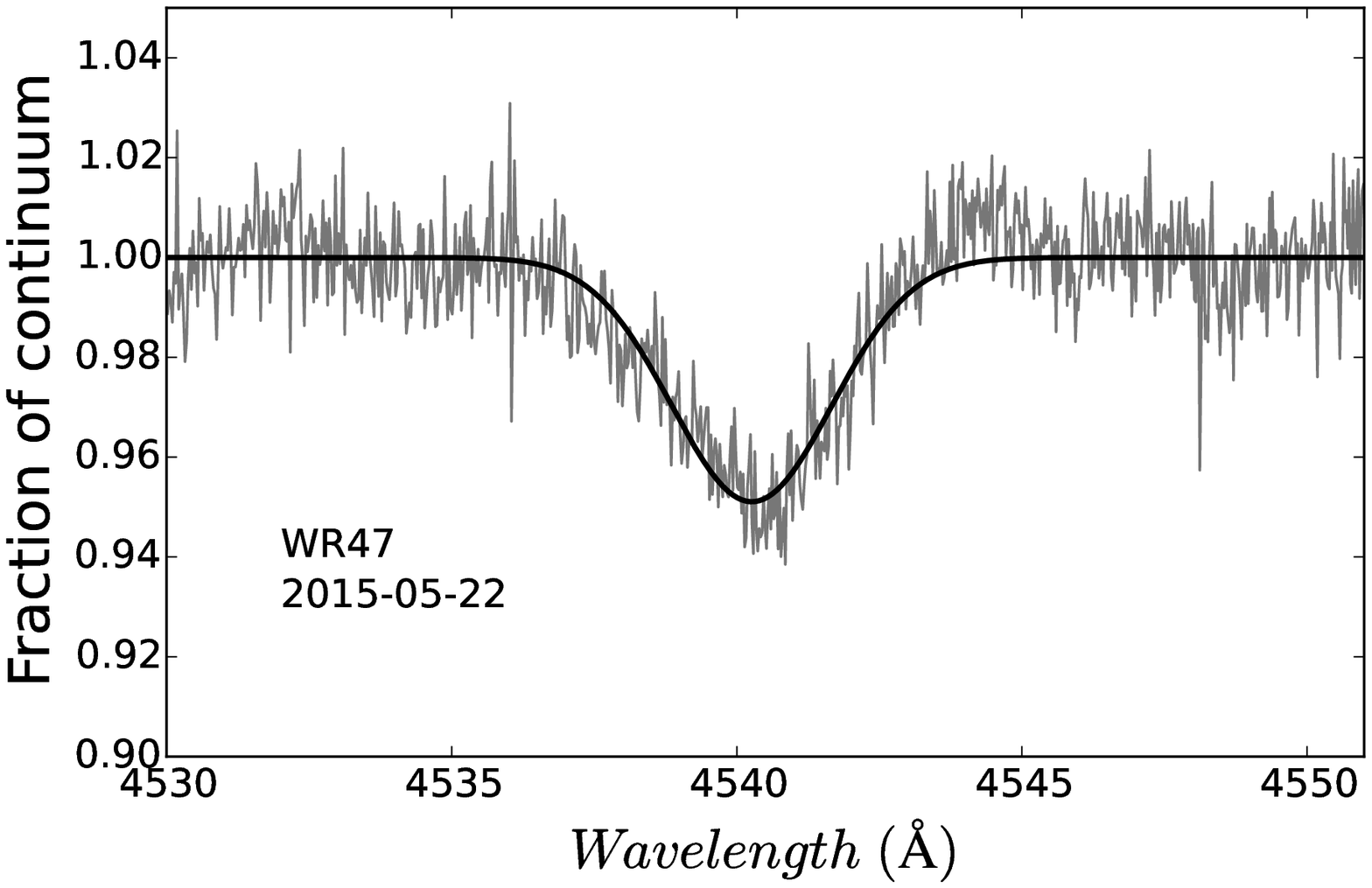}}
\centerline{\includegraphics[width=0.99\columnwidth]{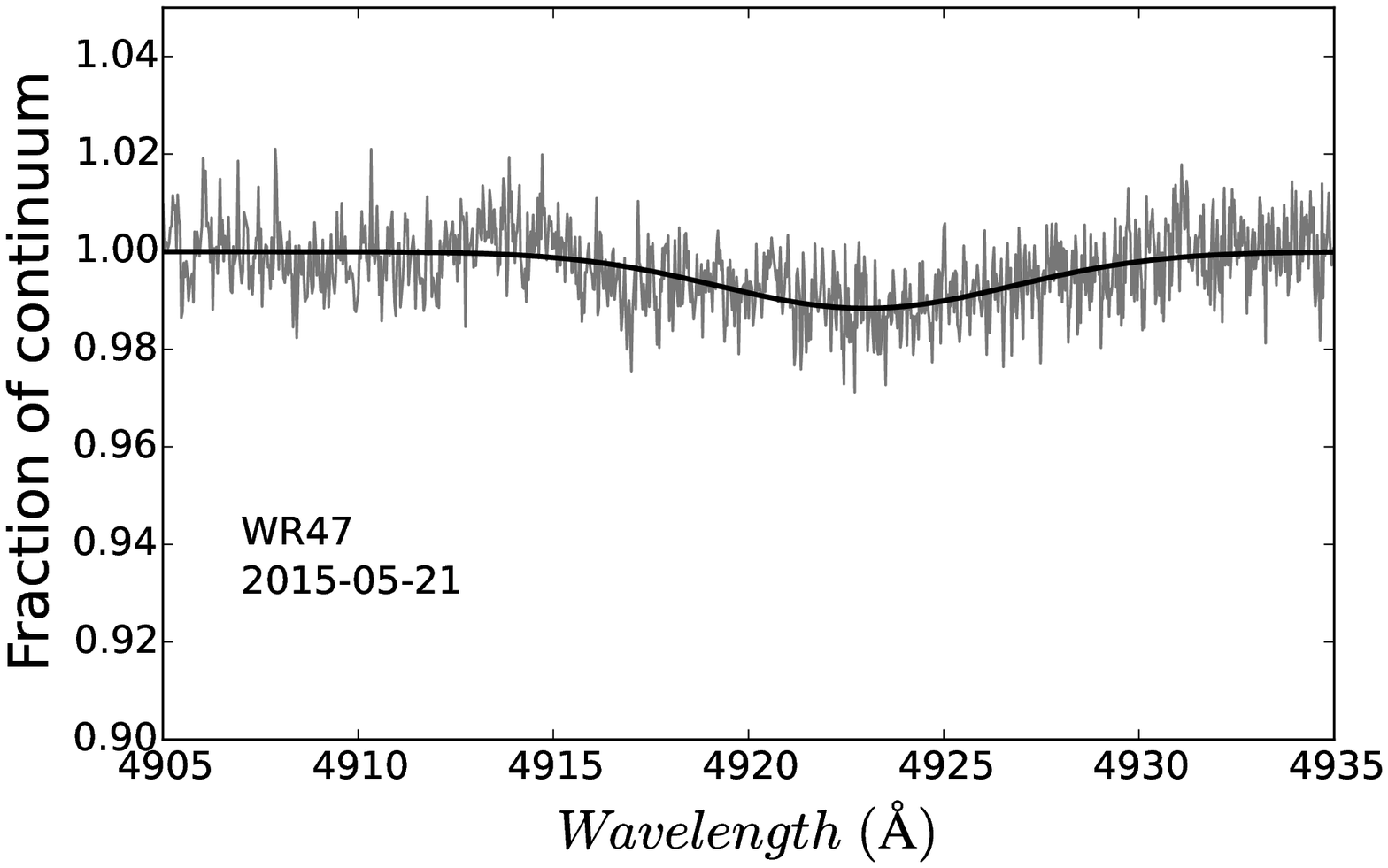}}
\caption{(Top) The HeII 4541 absorption line of WR 47 on 22 May 2015.
(Bottom) The HeI 4922 absorption line of WR 47 on 21 May 2015}\label{spectra}
\end{figure}

\begin{figure}
\centerline{\includegraphics[width=0.99\columnwidth]{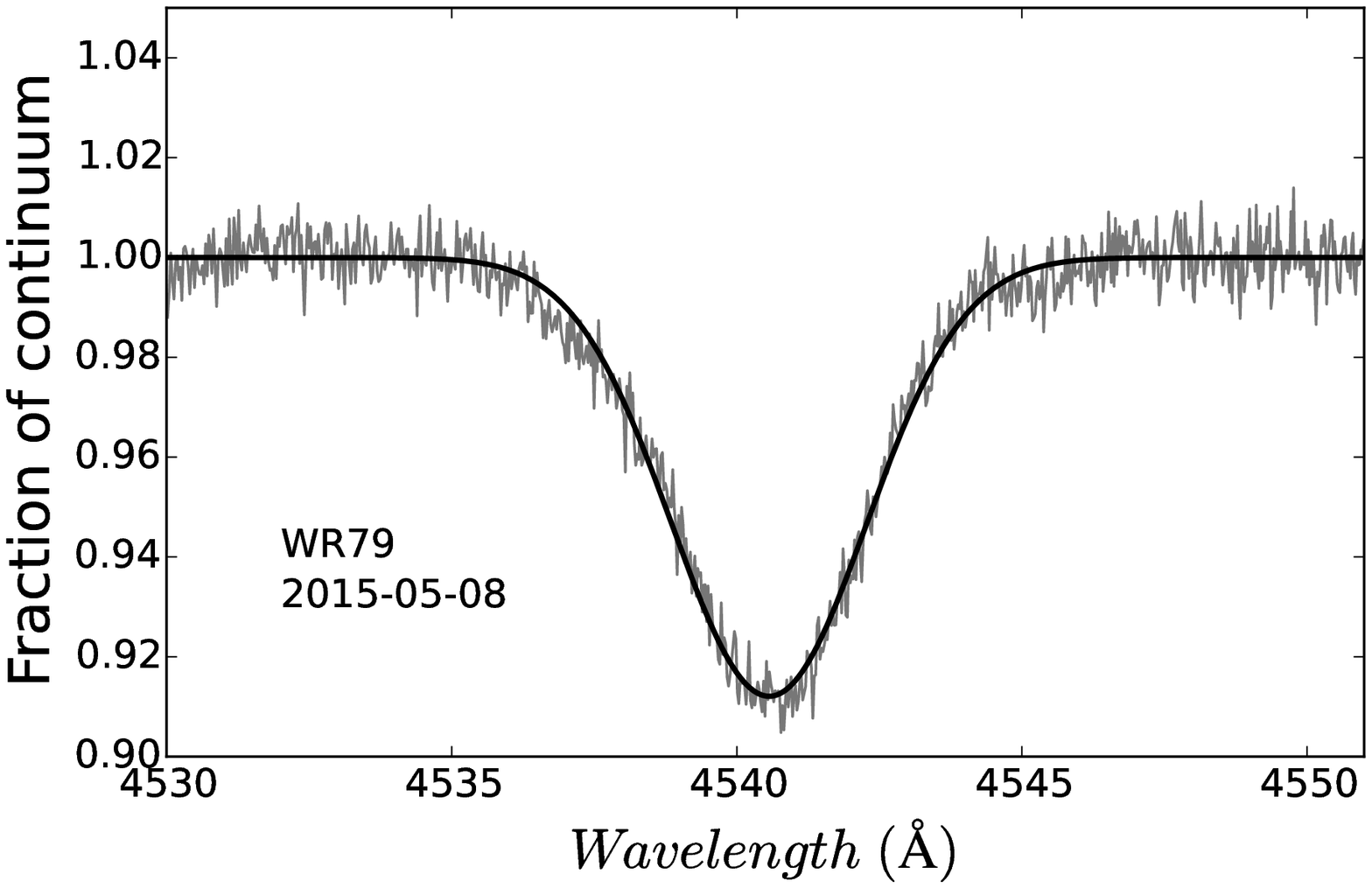}}
\centerline{\includegraphics[width=0.99\columnwidth]{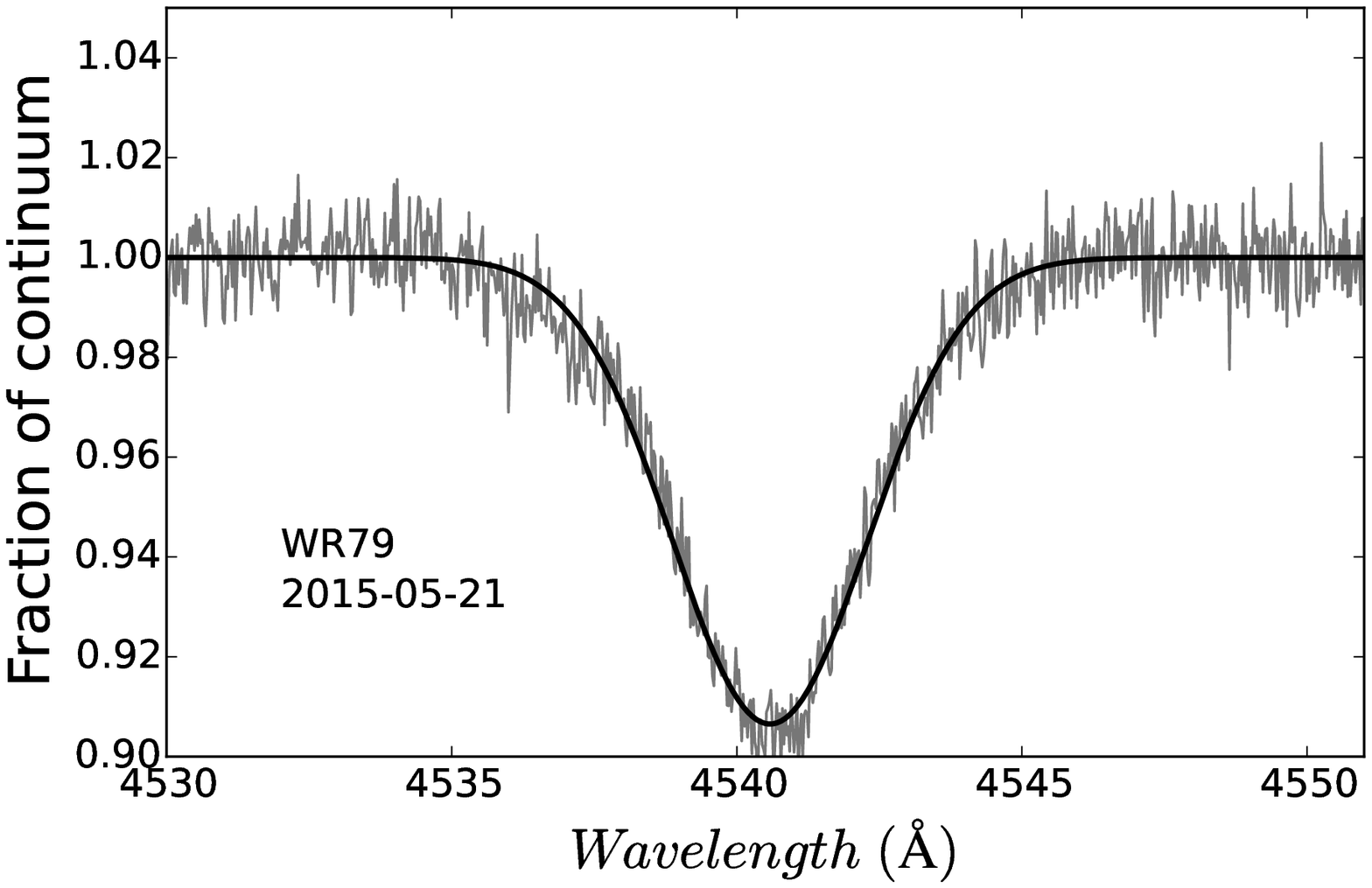}}
\centerline{\includegraphics[width=0.99\columnwidth]{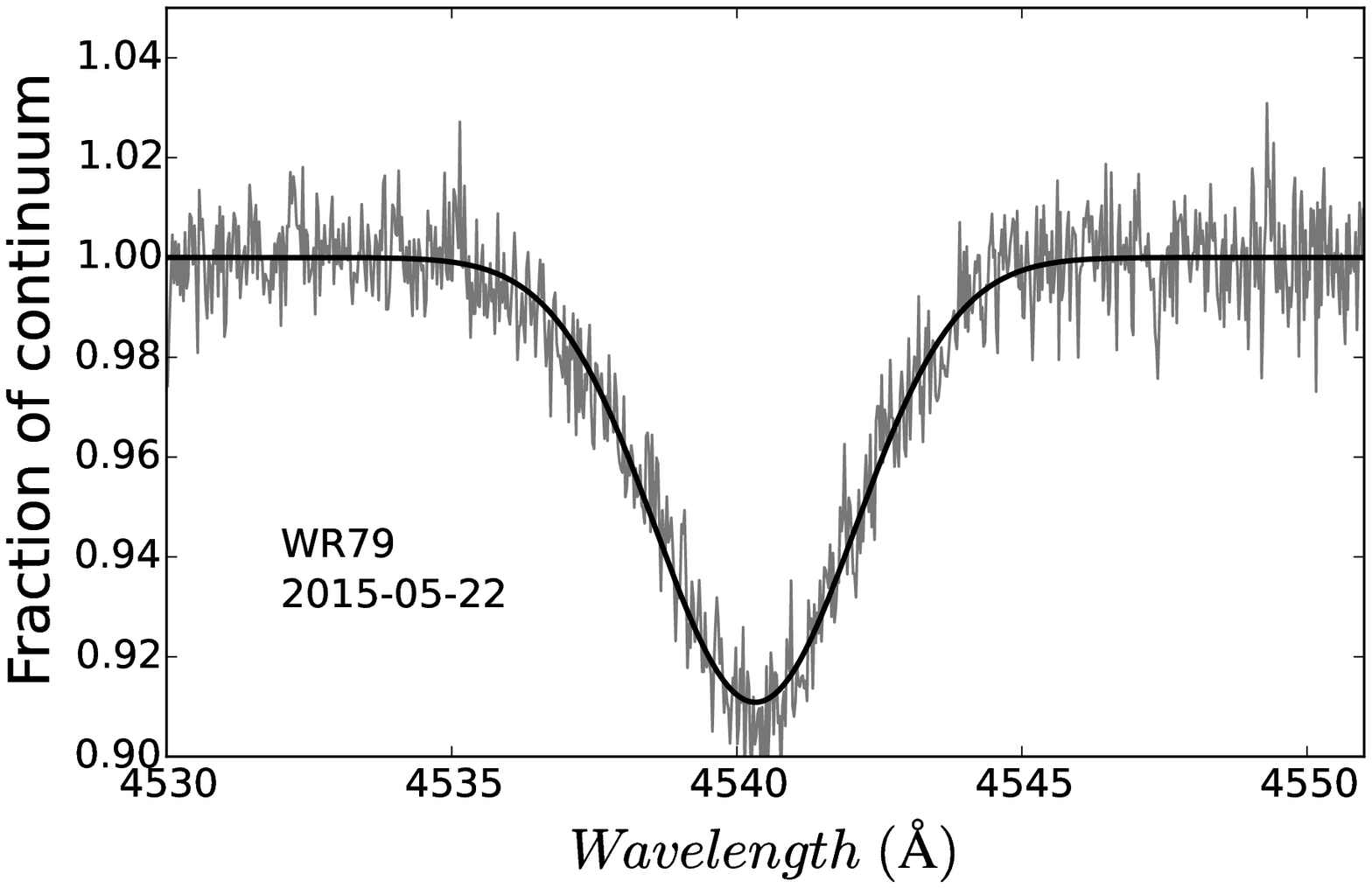}}
\caption{(Top) The HeII 4541 absorption line of WR 79 on 08 May 2015. 
(Middle) Same as above but on 21 May 2015.
(Bottom) Same as above but on 22 May 2015.}\label{spectra}
\end{figure}

\begin{figure}
\centerline{\includegraphics[width=0.99\columnwidth]{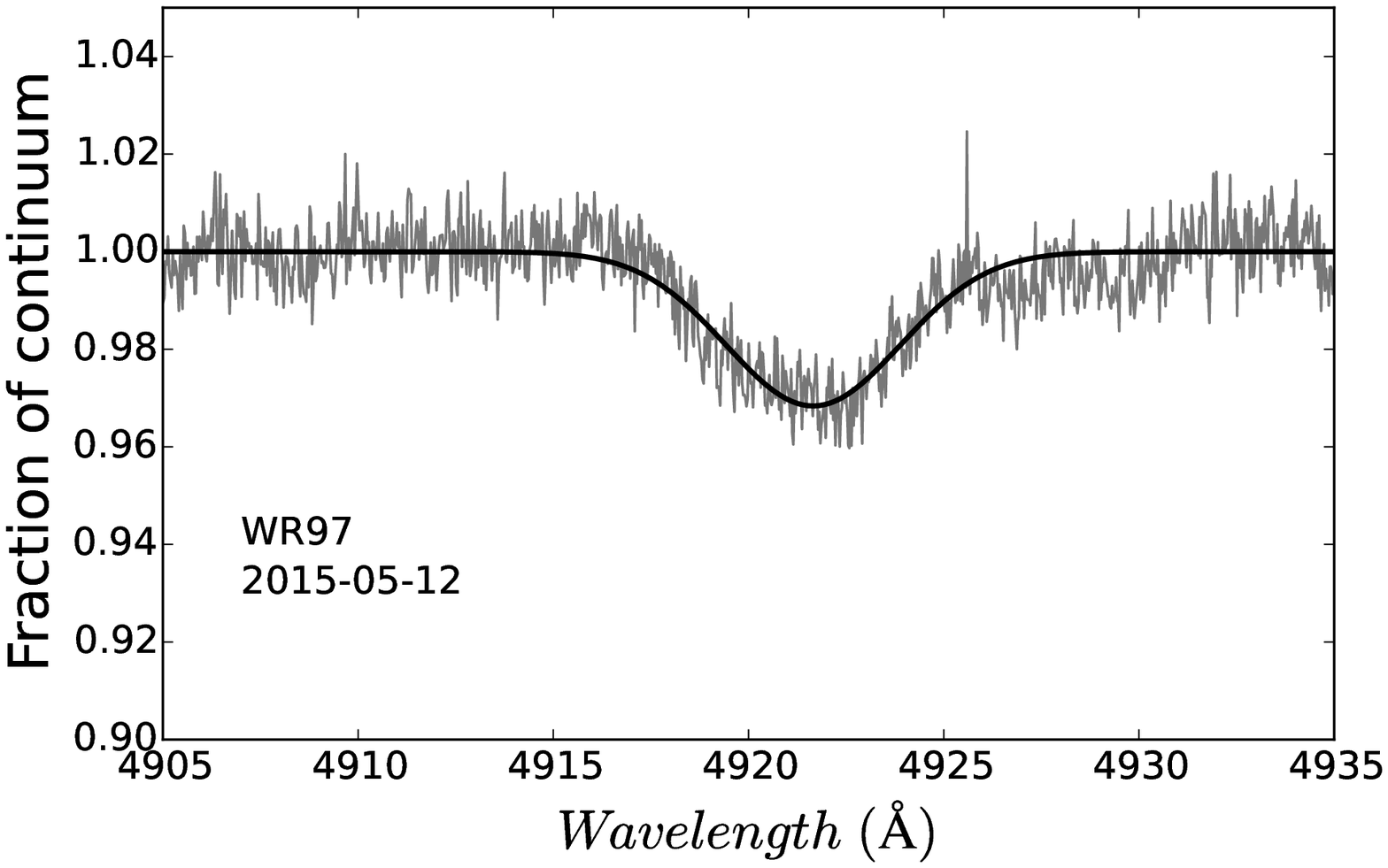}}
\centerline{\includegraphics[width=0.99\columnwidth]{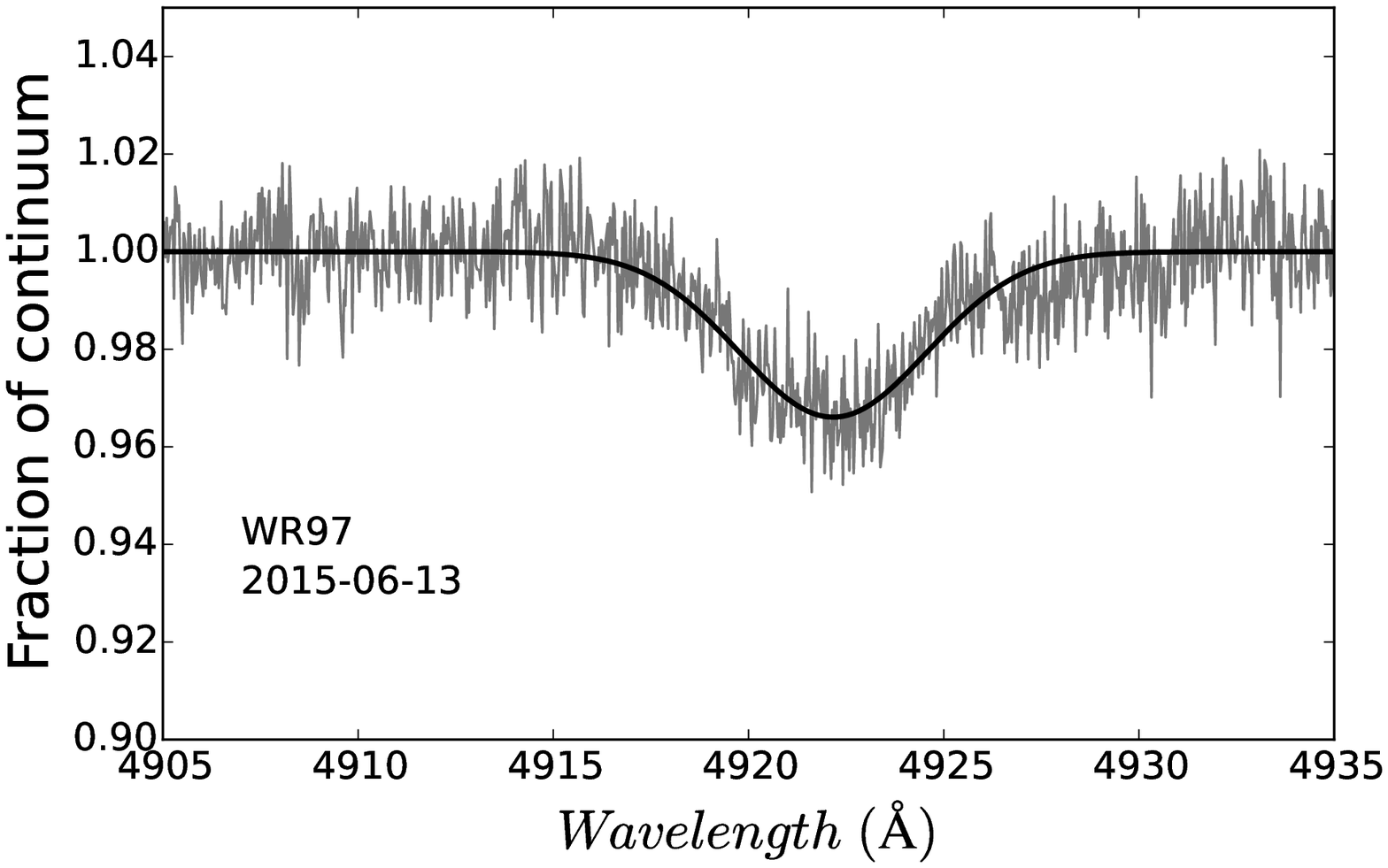}}
\centerline{\includegraphics[width=0.99\columnwidth]{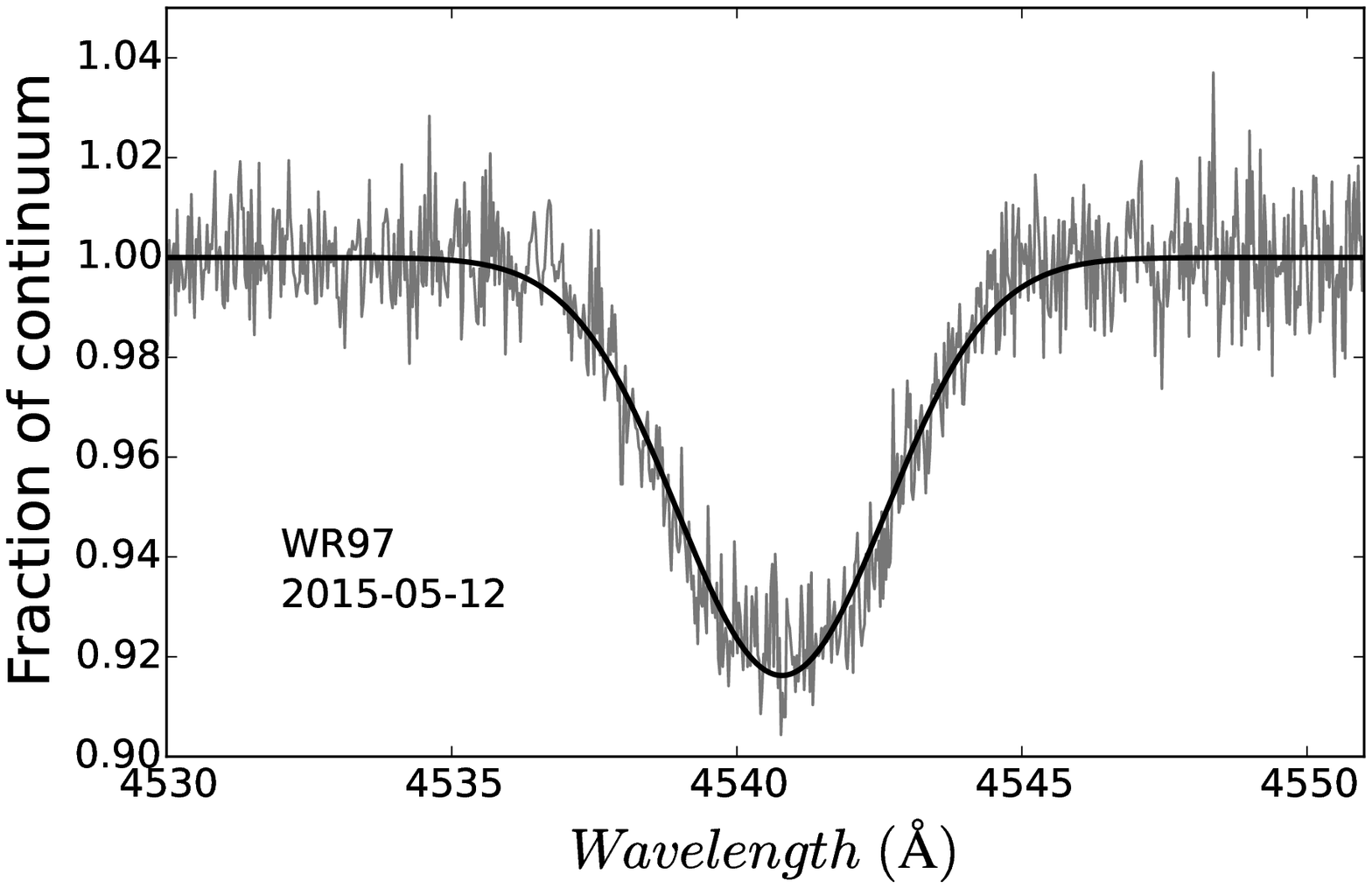}}
\centerline{\includegraphics[width=0.99\columnwidth]{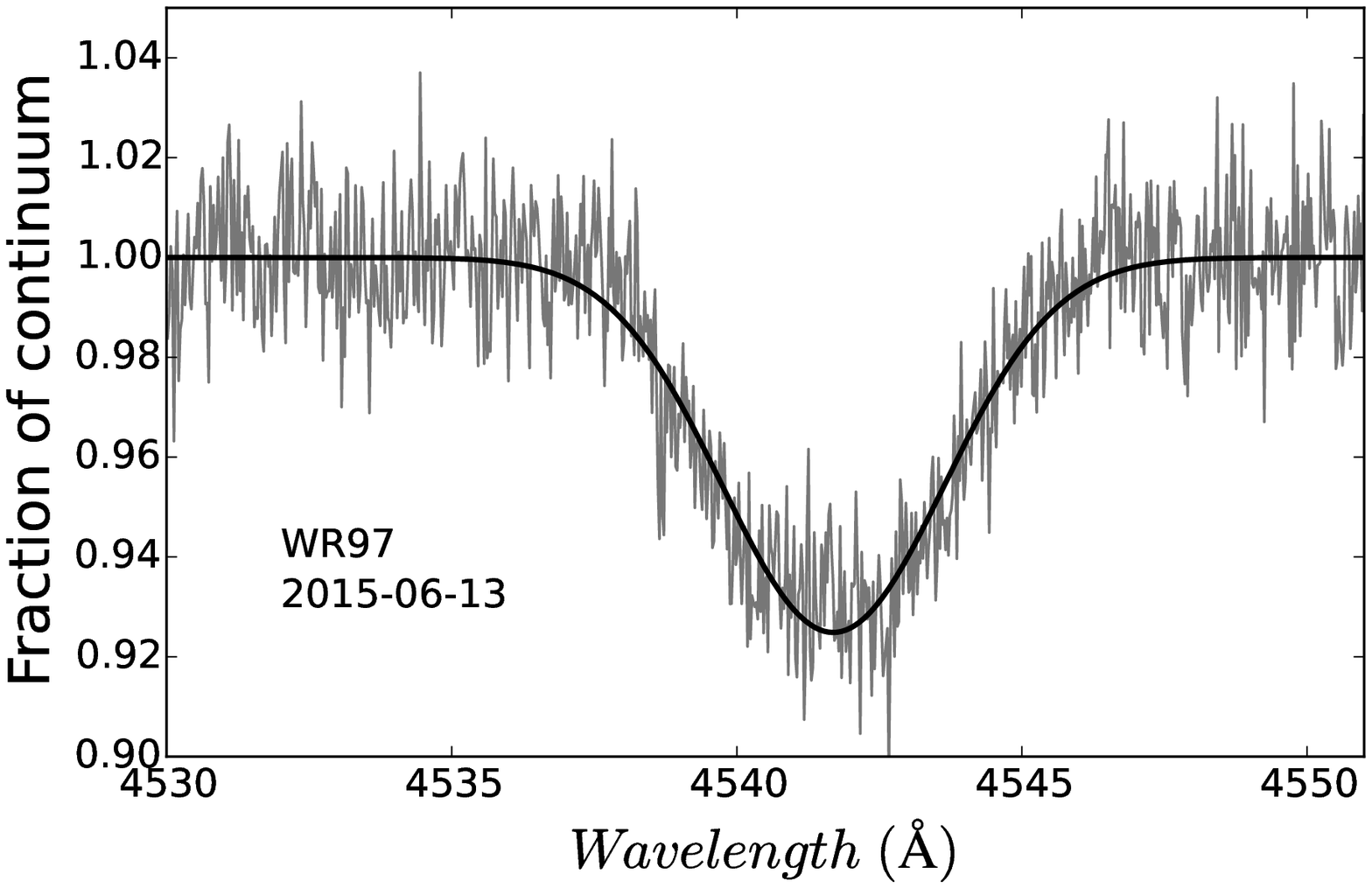}}

\caption{(Top) The HeI 4922 absorption line of WR 97 on 12 May 2015. 
(Second from Top) Same as above but on 13 June 2015.
(Third from Top) The HeII 4541 absorption line of WR 97 on 12 May 2015.
(Bottom) The HeII 4541 absorption line of WR 97 on 13 June 2015.}\label{spectra}
\end{figure}

\begin{figure}
\centerline{\includegraphics[width=0.99\columnwidth]{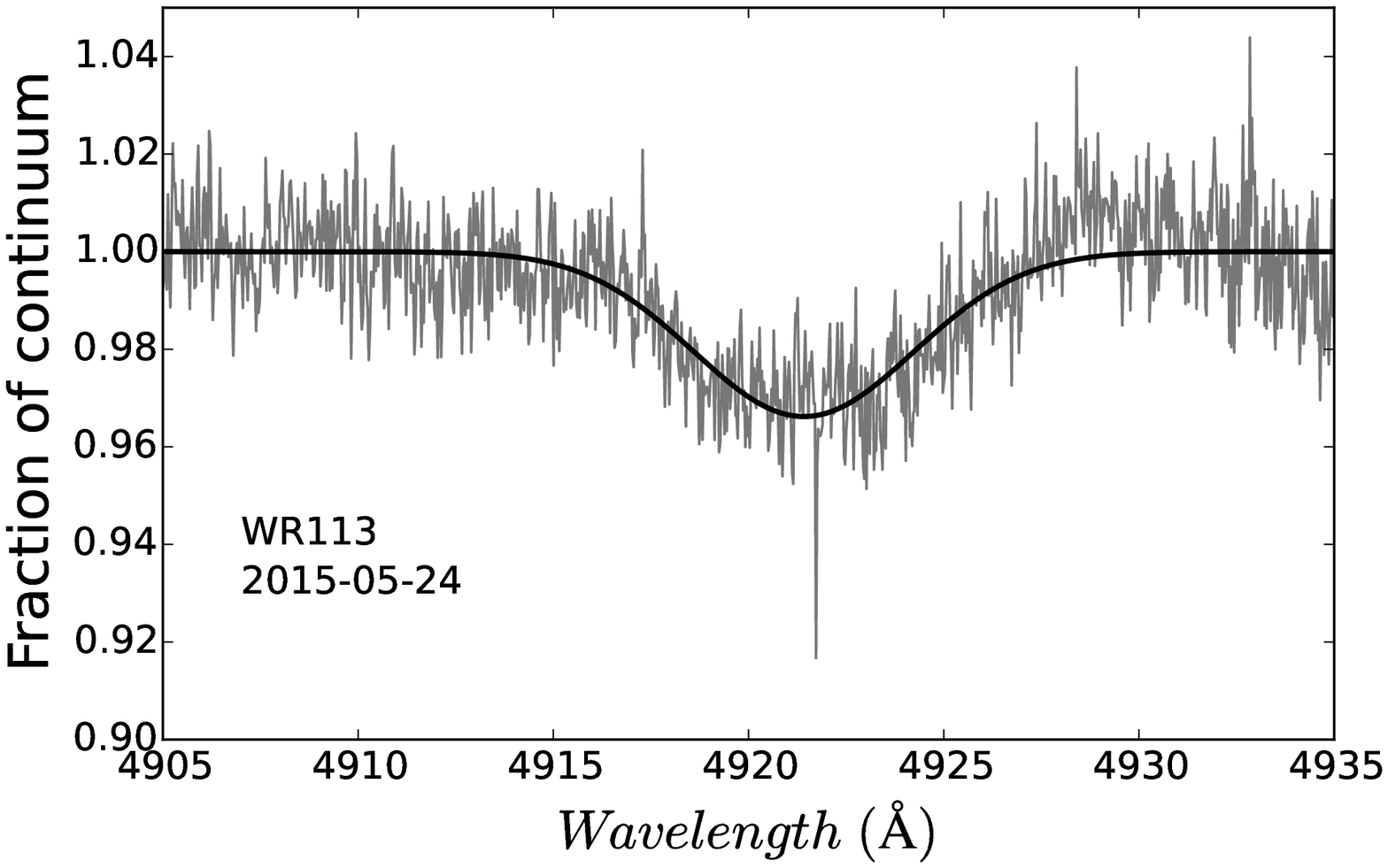}}
\centerline{\includegraphics[width=0.99\columnwidth]{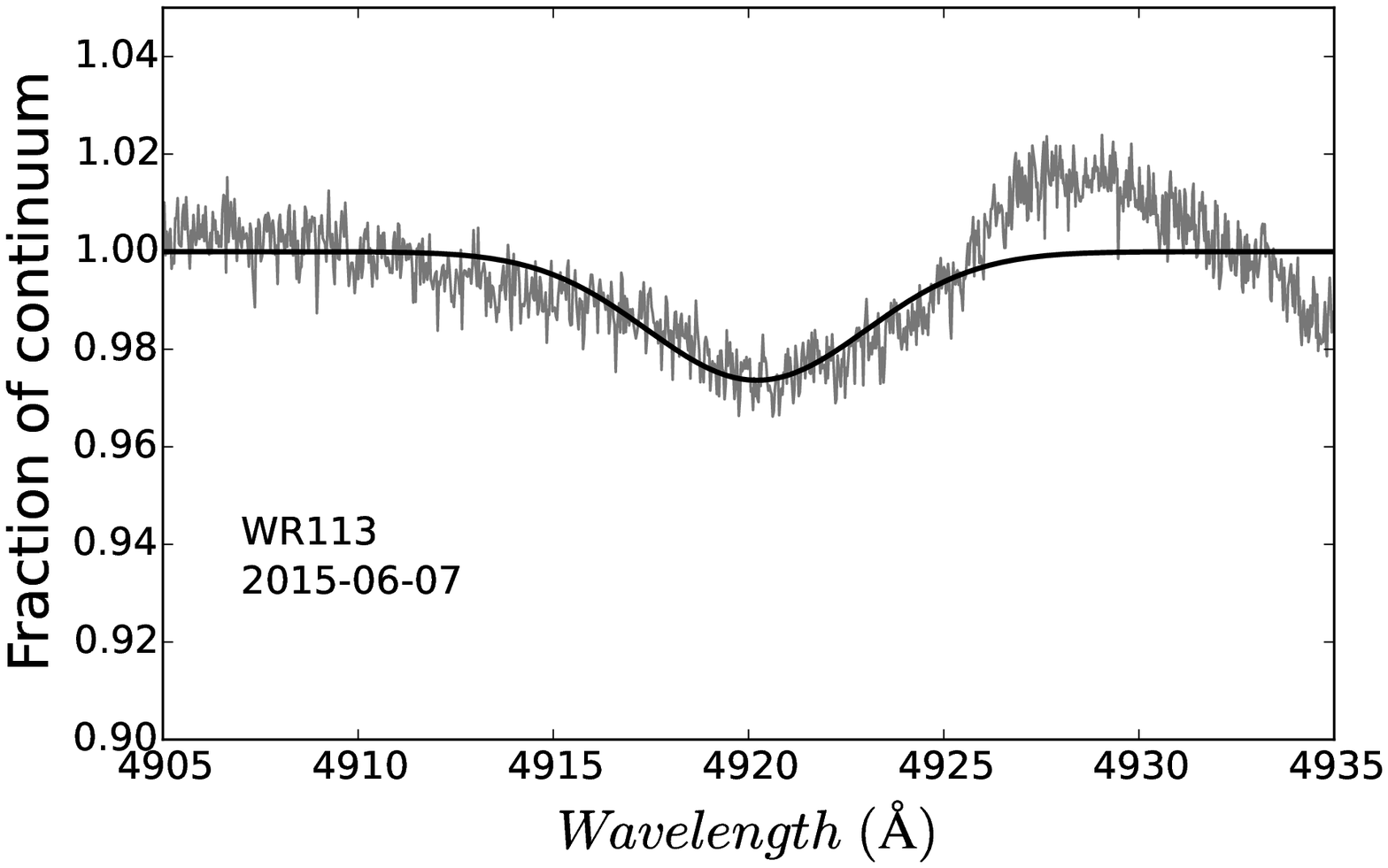}}
\centerline{\includegraphics[width=0.99\columnwidth]{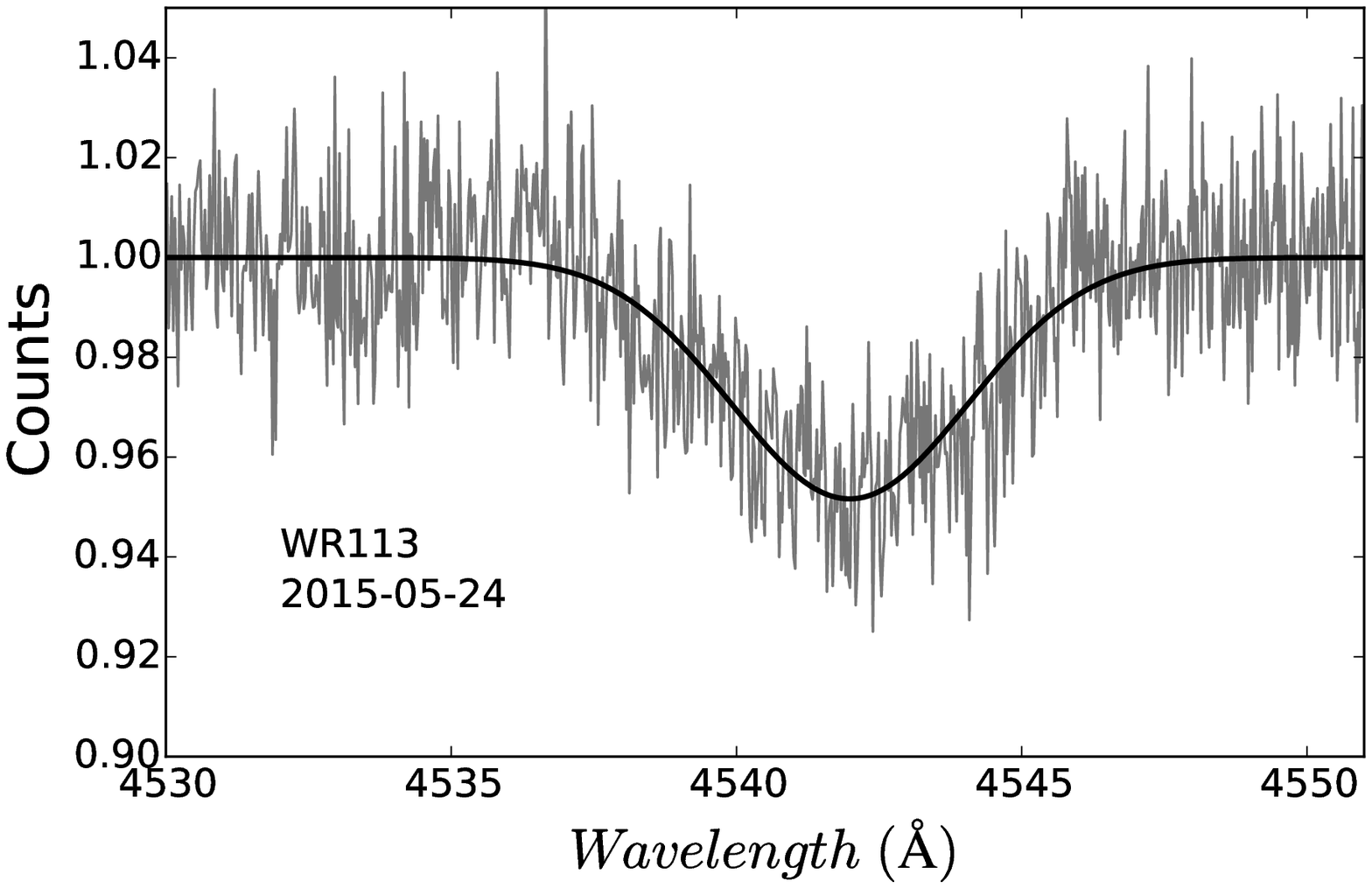}}
\centerline{\includegraphics[width=0.99\columnwidth]{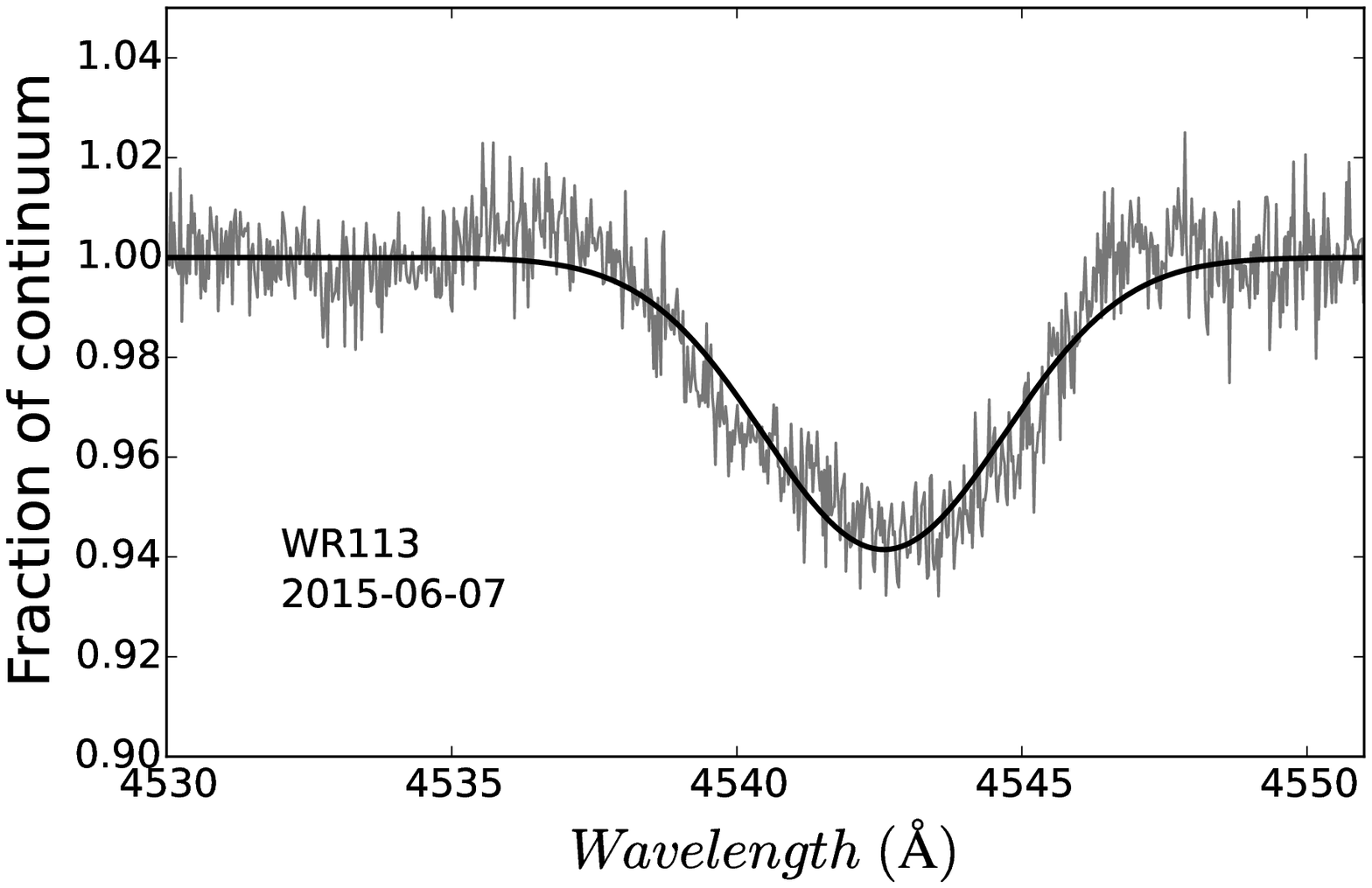}}
\caption{(Top) The HeI 4922 absorption line of WR 113 on 24 May 2015. 
(Second from Top) Same as above but on 07 June 2015.
(Third from Top) The HeII 4541 absorption line of WR 113 on 24 May 2015. 
(Bottom) The HeII 4541 absorption line of WR 113 on 07 June 2015. }\label{spectra}
\end{figure}

%\begin{figure}
%\centerline{\includegraphics[width=0.99\columnwidth]{fig10}}
%\caption{Observed HeI versus HeII $vsini$}\label{spectra}
%\end{figure}

\newpage

In Table 1, we report our measured HeI and HeII FWHM and velocities for each of
eight O stars in WR + O binaries, for $\tau$ Sco, and for WR 127 from Figure 8 of \citet{del11}. HeI absorption lines were 
detected in five O stars by SALT (WR 21, WR 31, WR 42, WR 97 and WR 113), and in WR 127 by \citet{del11},
while the  HeII $\lambda 4541$ line was detected in these same five stars as well as in
WR 30, WR 47 and WR 79.

The smallest FWHM measured in any of the target stars in our sample is $3.62\pm0.04 \AA$ (for the HeII $\lambda 4541$ line) which
corresponds to a $v.$sin i of $89\pm11 \ $km/s for WR 21. The largest FWHM measured is $7.25\pm0.05 \AA$ (for the HeI $\lambda 4922$ line), which
corresponded to a $v.$sin i of $315\pm21 \ $km/s for WR 42. For each star with more than one measurement of $v.$sin i,
(because of multiple epochs and/or more than one line measured), we adopt an average value from all its measurements. 
The average $v.$ sin i for our sample of eight O stars with measured HeII lines is 137 km/s. The average $v.$ sin i for our sample 
of six O stars with measured HeI lines is 268 km/s, which is nearly twice as large as the HeII lines'  $v.$ sin i. We return to this large difference below.

\begin{table*}
 \centering
  \caption{Measured FWHM and $v.$ sini of $\tau$ Sco and O stars in O+WR star binaries}\label{Tid}
  \begin{tabular}{@{}rllllclc@{}}
  \hline
Object & Observed\,&HeI $\lambda 4471, 4922$ FWHM (\AA)\,&HeI $v_e$ sin i (km/s)\,&He II $\lambda 4541$ FWHM (\AA)\,&HeII $\lambda 4541$ $v_e$ sin i (km/s)\\
 
\hline

WR21  & 2015-05-08 & $5.96 \pm 0.23$ & $246 \pm 20$ & $3.62 \pm 0.04$ & $89 \pm 11$  \\
WR21  & 2015-05-24 & $5.24 \pm 0.17$ & $209 \pm 16$ & $4.31 \pm 0.03$ & $120 \pm 16$ \\
WR30  & 2015-05-08 & ... & ... & $5.46 \pm 0.23$ & $178 \pm 27$ \\
WR31  & 2015-05-08 & $5.37 \pm 0.05$ & $216 \pm 15$ & $5.40 \pm 0.05$ & $176 \pm 23$ \\
WR42  & 2015-05-08 & $7.25 \pm 0.05$ & $315 \pm 21$ & $4.09 \pm 0.04$ & $109 \pm 14$ & \\
WR42  & 2015-05-21 & $...$ & $...$ & $4.36 \pm 0.05$ & $122 \pm 16$ & \\
WR47  & 2015-05-21 & ... & ... & $3.65 \pm 0.13$ & $90 \pm 13$  \\
WR47  & 2015-05-22 & ... & ... & $3.54 \pm 0.06$ & $86 \pm 11$  \\
WR79  & 2015-05-08 & ... & ... & $4.34 \pm 0.03$ & $121 \pm 16$ \\
WR79  & 2015-05-21 & ... & ... & $4.35 \pm 0.04$ & $121 \pm 16$ \\
WR79  & 2015-05-22 & ... & ... & $4.46 \pm 0.06$ & $127 \pm 17$ \\
WR97  & 2015-05-12 & $5.66 \pm 0.12$ & $230 \pm 17$ & $4.65 \pm 0.03$ & $136 \pm 18$  \\
WR97  & 2015-06-13 & $6.11 \pm 0.14$ & $254 \pm 18$ & $4.96 \pm 0.05$ & $152 \pm 20$ \\
WR113 & 2015-05-24 & $7.16 \pm 0.13$ & $310 \pm 22$ & $5.26 \pm 0.07$ & $168 \pm 22$ \\
WR113 & 2015-06-07 & $7.14 \pm 0.14$ & $309 \pm 22$ & $5.35 \pm 0.06$ & $172 \pm 23$ \\
WR127 & 2004 June-July &$7.0 \pm 0.5$ & $300 \pm 30$ & ...&... \\
\\
$\tau$ Sco& 2015-05-08 & $0.30 \pm 0.06$ & $24 \pm 3$ &...&... \\

\hline
\end{tabular}

\end{table*}

These data can now be combined with orbital inclinations gleaned from the literature. In table 2 we present these data, which yields (by far) the largest ensemble of rotational velocities of O stars in WR + O binaries yet assembled. The errors in derived $v_e$ are dominated by the uncertainties in orbital inclinations. In the final column of Table 2 we also list the synchronous rotation speeds and critical rotation speeds of the O-stars in each binary, using the already-measured masses and the spectral type-radius calibration of \citet{mar05}. A key result of this paper is that the average equatorial rotational velocity of six O stars, measured from HeI lines, is a highly super-synchronous 348 km/s. The average equatorial rotational velocity of eight O stars, measured from HeII lines, is 173 km/s, which is still significantly super-synchronous.

\begin{table*}
 \centering
  \caption{WR + O star binaries' properties }\label{Tid}
  \begin{tabular}{@{}rllllclc@{}}
 \hline
 System\,&Spectral Types\,&Period(d)\,&M(WR+O)$sin^3i (M_{\odot}$)\,& i(deg)\,&HeI $v_e$(km/s)\,&HeII $v_e$(km/s)\,&O-star v(sync)/v(crit)(km/s)\\

\hline
WR 21 & WN5o + O4-6 & 8.3 & 8.4 + 16.3 & $48-62$ & $278 - 331 $&$115 - 138 $&70/528 \\
WR 30 & WC6 + O6-8 & 18.8 & 15.4 + 31.9 &78 - 90 & ... & $178 - 182 $&25/818\\
WR 31 & WN4o + O8 V & 4.8 & 2.7 + 6.3 &40 - 62 & $244 - 336 $ & $199 - 274 $&89/382 \\
WR 42 & WC7 + O7 V & 7.9 & 3.7 + 6.2 & 38 - 44 & $453 - 511 $ & $157 - 177 $&66/361 \\
WR 47 & WN6o + O5 V & 6.2 & 40 + 47 & 67 - 90 & ... & $88 - 96 $&93/897 \\
WR 79 & WC7 + O5-8 & 8.9 & 1.8 + 4.9 & 34 - 45& ... & $174 - 220 $&53/321 \\
WR 97 & WN5b + O7 & 12.6 & 2.3 + 4.1 & 31 - 85 & $243 - 470 $ & $146 - 279  $&39/293 \\
WR 113 & WC8d + O8-9 IV & 29.7 & 10.6 + 22.3 & 70 - 90& $310 - 330  $& $170 -181 $&19/623 \\
WR 127 & WN5o + O8.5 V & 9.6 & 13.4 + 23.9 & 55 - 90 & $300 - 366 $ & ...&42/697 \\
\\
WR 11 & WC8 + O7.5 III & 78.5 & 6.8 + 21.6 & $63$& $220 $ & ... &9/545\\
WR 139 & WN5o + O6 III-V & 4.2 & 8.8 + 26.3 & $78.7$& $215 $ & ...&151/640\\
\hline

\end{tabular}\\
Spectral Types from \citet{cro15} and \citet{del11}); Mass functions and inclinations from \citet{lam96};\\ 
$v_e$ sin i references: This paper, except for WR 127 \citep{del11}, WR 11 \citep{baa90}, and WR 139 \citep{mar94}\\
\end{table*}
 
\section{Observations confront theory}

\subsection{HeI versus HeII lines' rotation speeds: oblate O stars}

\citet{ram13} have noted that values of $v.$ sin i for {\it single} O stars measured from HeII 
absorption lines are systematically lower than those measured from HeI lines by 25\% (see their Figure 10). They suggested that gravity darkening was responsible for this systematic difference. Rapid rotation produces equatorial centrifugal support of stars, hence oblateness and lower surface gravity at the equator, which is thus cooler and darker. Photospheric regions closer to the poles (equators) should then contribute more to the formation of He II (He I) lines.
Projected rotational velocities derived from He II lines should thus be lower than those from HeI lines, as observed in Figure 10 of \citet{ram13}.

von Zeipel's theorem \citep{vzp24} states that the radiative flux in a uniformly rotating star is proportional to the local effective gravity. Spectrographic observations, constrained by interferometric measurements of rapidly rotating stars show convincingly that latitudinal large temperature differences exist in the rapidly rotating stars Alpha Leo, Alpha Aquila and Achernar \citep{mca05, mon07, des14}. In particular, Alpha Leo displays equatorial (polar) temperatures of 10,314 K (15,400 K) \citep{mca05}.  Achernar similarly displays equatorial (polar) temperatures of 12,673 K (17,124 K) \citep{des14}. A recent review of the observations of these and other rapidly spinning stars is given by \citet{cla16}. These observations demonstrate conclusively that the polar temperatures of rapidly rotating stars can be 50\% hotter than their equatorial temperatures. 

Modern theoretical studies of gravity darkening \citep{lar11,cla16} conclude that von Zeipel's theorem is only applicable to slowly rotating stars. \citet{lar11} demonstrate that latitudinal variations in the effective temperature of a rapidly rotating star depend on the ratio of the equatorial velocity to the Keplerian velocity. Their model demonstrates good agreement with the above-noted observations of Alpha Leo and Alpha Aquila. \citet{cla16} obtains good agreement with observed gravity darkening indices for six rapidly rotating stars by considering optical depth effects in these stars' atmospheres. 

While our primary goal in undertaking this observational study was to determine whether RLOF-induced, super-synchronous rotation is seen in the O stars in WR + O binaries, an unexpected bonus has emerged. {\it Our observations indicate that not only are the O stars rotating super-synchronously, they also display large variations of effective temperature with latitude, suggesting that they may be oblate}. Since we are interested in the speeds of rotation at the equators of our O stars, it is clear that the velocities we derive from the HeI lines, and {\it not} those of the HeII lines are the speeds to use.

\subsection{Tides and RLOF}

Mass transfer during RLOF, wherein some of the mass lost by a donor star is accreted by its companion, is accompanied by angular momentum transfer, which forces the mass gainer to spin-up. It was demonstrated by \citet{pac81} that when the RLOF-process in a case B binary\footnote{A case B binary has an initial orbital period such that RLOF starts while the mass loser is hydrogen-shell burning, i.e. prior to core helium burning. The periods of such binaries range between about 10 days and 1000 days.} is quasi-conservative, then soon after the onset of this process the mass gainer is spun up to its critical Keplerian speed. Since the observed rotational velocity of the O-type companions in all 10 of the WR binaries in which it has now been measured is super-synchronous, it is tempting to conclude that mass transfer and spin-up have played important roles during the progenitor evolution. 

To further explore this scenario we assume that, very soon after the RLOF process began in the binaries we now observe, the O-type mass gainer was rotating critically. This corresponds to Keplerian rotation speeds of $\sim$ 530 km/s. The average rotation speed of six O-type companions with HeI lines in our observed binary sample is 348 km/s, close to 65\% of the Keplerian value. The evolutionary timescale of a WR star is typically of the order of a few hundred thousand years. Thus the 348 km/s observed rotation speed means that, even over this short timescale, tidal interactions must be efficient, able to cause significant spin-down of the O-type companions. 

We conclude by noting that the two longest period WR binaries in our sample (WR 113 at P= 29.7 d and WR 11 with P = 78.5 d) each have an O-type companion that is rotating super-synchronously, but with speeds that are significantly below critical. We are unaware of a tidal-synchronization theory (e.g., \citet{hut81,tas96,zan13}) that is capable of explaining a possible spin-down from critical rotation to the presently observed super-synchronous values in binaries with such large periods.

A detailed set of massive star binary evolution models, including RLOF angular momentum transfer and tidal braking by oblate stellar models, against which to compare our results, is lacking in the literature. Detailed models and population synthesis simulations confronting the observations reported here will be published in a separate paper \citep{van16}.    

\section{Conclusions}\label{conclusions}
WR + O binaries may be progenitors of ultra-luminous supernovae, long-duration gamma-ray bursts and BH-BH mergers.
Theory predicts that the O stars in WR + O binaries must have accreted significant amounts of angular momentum during RLOF from their companions. Only two O stars in WR+O binaries have previously published, measured values of $v.$ sin i, and one other has a rough estimate. We report new $v.$ sin i measurements for 8 O stars in WR + O binaries.  Using the literature values of i we then find the average equatorial rotational value of 348 km/s for 6 O stars, ranging from 237 to 482 km/s, from HeI lines. These values are highly super-synchronous. The observed super-synchronicity is in qualitative agreement with the predictions of short period, massive binary evolution models which include angular momentum transfer during RLOF and tidal braking afterwards. However, the super-synchronous nature of the two longest period binaries is a challenge to current tidal braking theory. We also find that the rotation rates of O stars in WR + O binaries, measured from HeI lines, are on average twice as large as those measured from HeII lines. We conclude that we are observing gravity darkening in these O stars, and predict that they are oblate.
%==================================================================
\section*{Acknowledgments}

All of the new observations reported in this paper were obtained with the High Resolution Spectrograph (HRS) of the Southern African Large Telescope (SALT).
We gratefully acknowledge the fine support of the astronomers and operators at the SALT Observatory. The generosity of the late Paul Newman and the Newman Foundation has made AMNH's participation in SALT possible; MMS gratefully acknowledges that support. S.M.C. acknowledges the South African Astronomical Observatory and the National Research Foundation of South Africa for support during this project. We thank Ray Sharples and the Durham University team for construction and delivery of an excellent High Resolution Spectrograph. This research made use of Astropy, a community-developed core Python package for Astronomy (Astropy Collaboration, 2013). We thank Nobert Langer and Yong Shao for helpful suggestions. We also thank Professor Ian Howarth for two careful and thoughtful referee's reports, which resulted in significant improvements to the paper.

%==================================================================

\newpage

\label{lastpage}

\end{document}